\documentclass[12pt,english,super,sort&compress]{elsarticle}
\usepackage{helvet}

\usepackage[T1]{fontenc}
\usepackage[latin9]{inputenc}
\usepackage{geometry}
\usepackage{color}
\usepackage{amsmath}
\usepackage{amssymb}
\usepackage{graphicx}

\makeatletter
\usepackage{babel}

\usepackage{epstopdf}

    \def\ps@pprintTitle{%
       \let\@oddhead\@empty
       \let\@evenhead\@empty
       \def\@oddfoot{\reset@font\hfil\thepage\hfil}
       \let\@evenfoot\@oddfoot
    }

\@ifundefined{showcaptionsetup}{}{%
 \PassOptionsToPackage{caption=false}{subfig}}
\usepackage{subfig}
\makeatother

\usepackage{babel}
\begin{document}

\begin{frontmatter}{}

\title{A Multi-Scale Approach to Describe Electrical Impulses Propagating
along Actin Filaments in both Intracellular and In-vitro Conditions}

\author{Christian Hunley, Diego Uribe, and Marcelo Marucho\footnote{Email:marcelo.marucho@utsa.edu}}

\address{Department of Physics and Astronomy, The University of Texas at San
Antonio, San Antonio, TX 78249-5003}
\begin{abstract}
An accurate and efficient characterization of the polyelectrolyte
properties for cytoskeleton filaments are key to the molecular understanding
of electrical signal propagation, bundle and network formation, as
well as other relevant physicochemical processes associated with biological
functions in eukaryotic cells and their potential nanotechnological
applications. In this article, we introduce an innovative multi-scale
approach able to account for the atomistic details of a proteins molecular
structure, its biological environment, and their impact on electrical
impulses propagating along wild type Actin filaments. The approach
provides a novel, simple, accurate, approximate analytic expression
for the characterization of electrical impulses in the shape of soliton
waveforms. It has been successfully used to determine the effects
of electrolyte conditions and voltage stimulus on the electrical impulse
shape, attenuation and kern propagation velocity in these systems.
It has been shown that the formulation is capable of accounting for
the details of electro-osmosis and convection, as well as the electrical
double layer of G-actins includding the electrical conductivity and
capacitance. The approach predicts higher electrical conductivity,
linear capacitance and nonlinear accumulation of charge in intracellular
condictions. Our results also show a siginificant influence of the
voltage input on the electrical impulse shape, attenuation and propagation
velocity. The filament is able to sustain the soliton propagation
at almost constant kern velocity in the in-vitro condition, but the
soliton displays a remarkable deceleration in the intracellular condition.
Additionally, the solitons are narrower and travel faster at higher
voltage input. Whereas, the voltage input does not play an important
role on the soliton kern velocity in the in-vitro condition. Overall,
Our results predict the propagation of electrical signal impulses
in the form of solitons for the range of voltage stimulus and electrolyte
solutions typically present in intracellular and in-vitro conditions.
This multi-scale theory may also be applicable to other highly charged
rod-like polyelectrolytes with relevance in biomedicine and biophysics.
It is also able to account for molecular structure conformation (mutation)
and biological environment (protonations/deprotonations) changes often
present in pathological conditions.
\end{abstract}

\end{frontmatter}{}

\begin{keywords} Electrical signal propagation, Actin filaments,
Theoretical developments, Electrochemistry, Polyelectrolytes \end{keywords}

\section{Introduction}

Actin filaments (F-actin) are long charged rod-like cytoskeleton polymers,
which carry out many important biological activities in eukaryotic
cells.\cite{dos2008actin,woolf2009nanoneuroscience} These microfilaments
have recently gained wide notoriety for their fascinating polyelectrolyte
properties.\cite{Janmey2014} According to single filament experiments
in solution,\cite{Cantiello1991OSMOTICALLY,actin_like_cable_cantiello_1993}
F-actin have been shown to sustain ionic conductance and transmit
electrical currents in the form of localized counterionic waves about
the polymer's surface. The velocity of propagation along the surface
of an actin microfilament is consistent with the velocity of propagation
for neuronal impulses. Hence, in principle, concurrent propagation
of both electrical signals along actin microfilaments, and electrochemical
currents along the axonal membrane are highly possible. Another intriguing
property of these proteins is their capacity for overcoming electrostatic
interactions to form higher-order structures (bundles and networks)
in the cytoplasm. For instance, cytoskeletal filaments are often directly
connected with both ionotropic and metabotropic types of membrane
embedded receptors, thereby linking synaptic inputs to intracellular
functions. \cite{Goldman1952} Conducting microfilaments may also
govern at least some aspects of overall ion channel behavior within
microvilli. \cite{LANGE2000561,GartzkeC1333} 

All of these observations provide strong evidence on the polyelectrolyte
nature of F-actin, which provides unique, yet still poorly understood,
conducting and bundling formation properties in a variety of neuron
activities including intracellular information processing, regulating
developmental plasticity, and mediating transport. Certainly, the
molecular understanding of the polyelectrolyte properties of cytoskeleton
filaments will not only open unexplored frontiers in biology and biomedicine,
but it will also be crucial for the development of reliable, highly
functioning small devices with biotechnological applications such
as bionanosensors and computing bionanoprocessors. \cite{sundberg2006actin,Arsenault2007,galland2013,Patolsky2004,Korten2010}
Therefore, it is of crucial importance to determine the underlying
biophysical principles and molecular mechanisms that support the ionic
conductance and electrical impulse transmission in Actin filaments
under a variety of biological environments. 

\begin{figure}
\centering{}\includegraphics[scale=0.07]{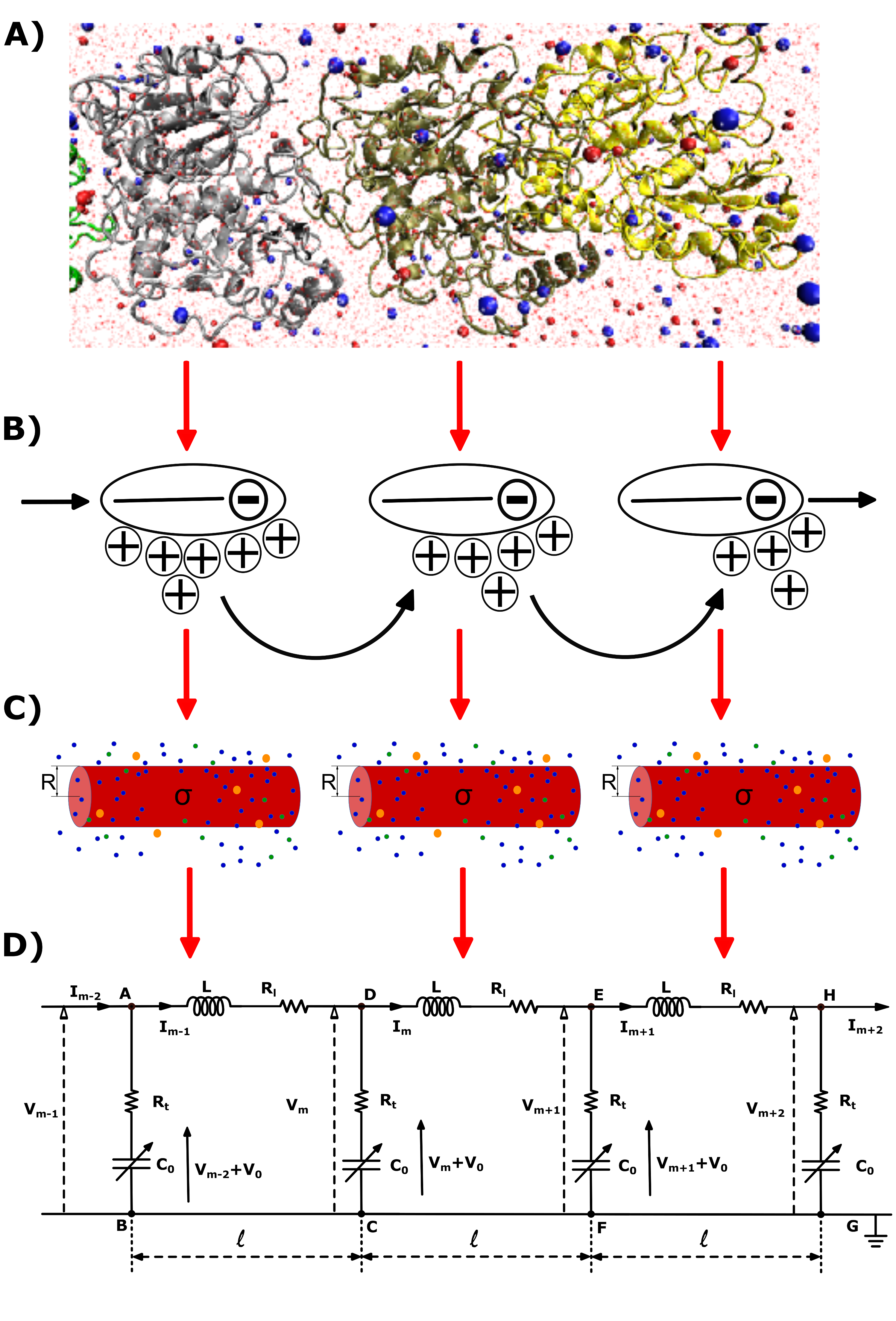}\caption{A) Molecular Structure Model. B) Ion Condensation Theory. C) Cylindrical
Biomolecule Model D) Dispersive Transmission Line Model\label{fig_description_ions_move}}
\end{figure}

The current understanding of these phenomena builds on a pioneers
work, Fumio Oosawa suggested around 50 years ago that electric signals
could be channeled through a medium along a microfilament due to the
electrolyte solution forming an electron cloud along the filament
length \cite{Oosawa_1970}. It was Manning who introduced condensation
theory \cite{manning_1978}, which provided the foundation for linear
polymers to enable electrical currents in the form of ionic movements.
As charged polyelectrolytes, cytoskeleton filaments may contain a
proportion of their surrounding counterions in the form of a dense
or \textquotedbl{}condensed\textquotedbl{} cloud about their surface,
as long as, there is a sufficiently high linear charge density, a
critical concentration of multivalent ions, and a small dielectric
constant of the surrounding medium. These criteria are indeed met
for actin filaments in neurons. \cite{Manning_1969,Oosawa_1970,Zimm_1986}
Further, molecular structure analysis indicates the distribution of
counterion clouds is nonuniform along the filament\textquoteright s
length. This is because F-actin originates from the linear polymerization
of globular actin (G-actin) units. Each of these units have tight
binding sites that mediate head-to-tail interactions to form a double-stranded
helix. Therefore, resembling a solenoid with a fluctuating current
flowing as a result of voltage differences generated by the ends of
the filament. Additionally, with the filament core separated from
the rest of the ions in the bulk solution by the counterion condensation
cloud, this overcast of counterions may act as a dielectric medium
between the filament and bulk layer. Hence, providing F-actin both
resistive and capacitive behaviors that may be associated with a highly
conductive medium. This conduction along microfilaments is characterized
by the decomposition of an electrical input pulse into discrete delayed
charge portions. These patterns clearly indicate the existence of
charge centers with corresponding counter ion clouds along the polymer
axis. As a consequence, electrically forced ions entering one end
of the biopolymer will result in ions exiting the other end (Fig.
\ref{fig_description_ions_move}). Therefore, actin polymers may serve
as \textquotedbl{}electrical nanobiowires\textquotedbl{} whom can
be modeled as nonlinear inhomogeneous transmission lines known to
propagate nonlinear dispersive solitary waves .\cite{kolosick1974properties}
These waves can take the form of localized electrical signal impulses
.\cite{Noguchi1974,Lonngren1978127,Novikov1984} However, this basic
understanding on electrical impulses propagating along actin filaments
does not account for all conductance properties of microfilament bundles.
\cite{Bret_Le_Zimm_1984} More recent approaches, based on Gouy-Chapman
electrical double layer type models and mean-field Poisson-Boltzmann
theories, provide further insight into the ionic equilibrium distributions
and electric potential properties near the polymer surface, which
arise from the charged polyelectrolyte surface, continuum solvent
dielectric medium, and mean electrostatic potential energy generated
by mixed salts comprised of point-like ions \cite{Newmancha1,Tuszynski2004,Sataric2009,Sekulic2011,Sekulic_MT_2015}.
These methods are shown to break down for cytoskeleton filaments under
certain situations, because they entail several approximations in
their treatment of the ions and solvent molecules. They don't account
for water crowding, ion size asymmetry or electrostatic ion correlation
effects, all of which are likely to play a fundamental role in providing
a quantitative description of the polyelectrolyte nature of cytoskeleton
filaments, and consequently, their conducting and electrical signal
propagation properties \cite{Kornyshev2007,JIANG2011153,Lamperski2015,Warshavsky2016}.

In this article, we introduce a novel multi-scale ($\textrm{atomic}(\textrm{Å})\rightarrow\textrm{monomer}(nm)\rightarrow\textrm{filament}(\mu m)$)
approach to describe nonlinear dispersive electrical impulses propagating
along Actin filaments (see Fig. 1). In section II, an atomistic model
for the F-actin \cite{CONG2008331} and its biological environment
is used to determine the polyelectrolyte properties and molecular
mechanisms governing G-actins in the polymerization state. This approach,
along with a suitable modification of Nernst-Planck theory \cite{Newmancha9},
are used to calculate the monomeric radial and axial flow resistances.
In addition, a more sophisticated approach based on a classical solvation
density functional theory \cite{Marucho2014,medasani2014ionic,Hunley2017,Marucho2016,Warshavsky2016}
is required to calculate the monomeric radial ionic capacitance. In
section III we use this monomer characterization to capture the biophysics
and biochemistry at larger (microfilament) scale distances. We utilize
those parameters in a nonlinear inhomogeneous transmission line prototype
model which accounts for the monomer-monomer interactions, and consequently,
the electrical impulse propagation along the filament length. A novel
approximate analytic solution is obtained for this model and utilized
in section IV to characterize the electrical impulse peak, width,
and velocity of propagation under several voltage stimulus and electrolyte
conditions. 

\section{G-actin Characterization in the Polymerization State}

\subsection{Cylindrical Biomolecule Model for G-actins}

\begin{figure}
\begin{centering}
\includegraphics[scale=0.05]{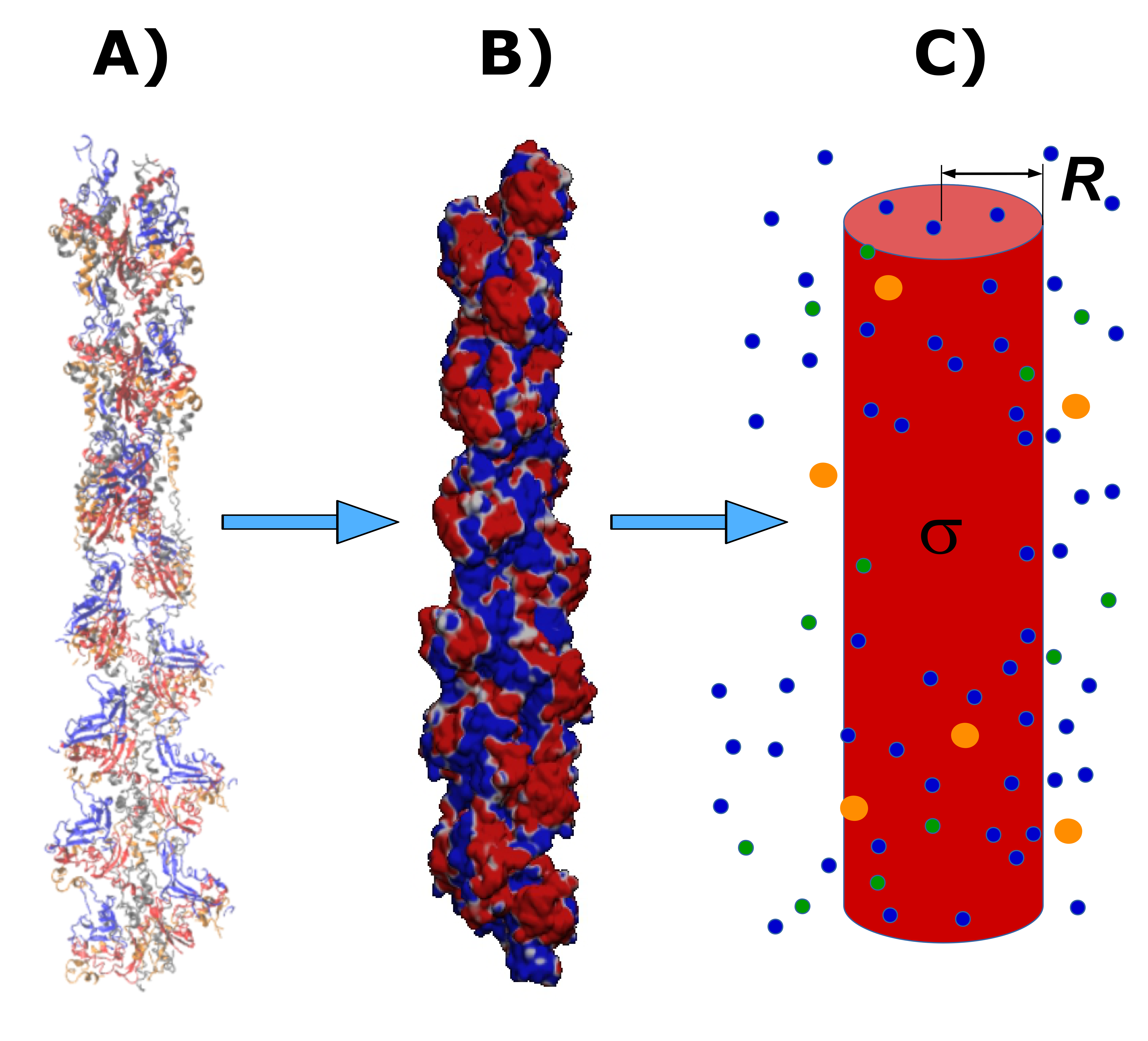}
\par\end{centering}
\caption{A) Cong Model B) Filament Volume C) Cylindrical Characterization\label{fig:Cong-Model} }
\end{figure}

The physicochemical properties of each monomer composing the actin
filament are different from those as single globular actin proteins,
because polymerization into filamentous form generates several conformational
changes on each monomer. Therefore, we retrieve the information on
the monomer molecular structure from one of the most recent 13 monomers,
biologically assembled wild type F-actin filament models posted on
the protein data bank: the Cong model \cite{CONG2008331} (see Fig.\ref{fig:Cong-Model}a).
It provides a detailed molecular characterization including the amino
acids sequence and the number and type of residues exposed to the
electrolyte. This uncharged molecular structure in pdb format is uploaded
into pdb2pqr webserver \cite{Dolinsky2004} to assign atomic charges
and sizes, add hydrogens, optimize the hydrogen bonding network, and
renormalize atomic charges of the residues exposed to the surface
due to pH effects (protonation/deprotonation process). The resulting
charged molecular structure at $pH=7.2$ is used to extract information
on the effective filament $L=Z_{max}-Z_{min}=422.20\mathring{A}$
and monomer length $\ell=L/13=5.4nm$, as well as the total filament
charge $Q=\sum q_{i}=-154e$, where $e$ is the electronic charge
(see Fig. \ref{fig:Cong-Model}b). The filament length and total charge
are used to estimate the filament linear charge density $\lambda=Q/L=-0.365\frac{e}{\mathring{A}}$.
The resulting structure is uploaded into \textquotedblleft 3v: voss
volume voxelator\textquotedblright{} webserver \cite{3vvoss2010}
to estimate the total filament volume $V_{p}=753005\mathring{A}^{3}$.
From here the effective monomer radius $R$ of the molecular structure
model is calculated $R=\sqrt{V_{p}/\left(2L\right)}=23.83\mathring{A}$.
The linear charge density and radius are subsequently used to calculate
the filament (= monomer) surface charge density $\sigma=\lambda/\left(2\pi R\right)=-0.039027\frac{C}{m^{2}}$.
This effective radius $R$, length $\ell$ and surface charge density
$\sigma$ are used in the next sections to determine the longitudinal
and transversal ionic flow resistances, capacitance and self-inductance
for each monomer along the filament length ((see Fig. \ref{fig:Cong-Model}c).

\subsection{Electrical and Conductive Properties of G-actins in Solutions}

In this approach, the electrical properties of a single G-actin are
characterized by a capacitor, two resistances and a self-inductance
component. The capacitor, whose capacitance changes with applied voltage,
originates the non-linearity behavior of the electrical impulse. The
structural periodicity in the arrangement of monomers generates the
dispersion of the electrical impulse along the filament. Whereas,
the losses in the transmission media is accounted for by a series
and shunt resistors, which represent the finite conductivity of the
conductors and the dielectric insulator between the conductors, respectively.
Additionally, the inductive component to the electrical properties
of the electrical impulse is due to F-actin\textquoteright s double-stranded
helical structure, which induces the ionic flow in a solenoidal manner
around each monomers. A detailed characterization of these properties
is provided below.

\subsubsection{Longitudinal and Transversal Ionic Flow Resistances}

We use transport and Ohm's laws as well as Navier-Stokes and Poisson's
theories to obtain simple and accurate, approximate analytic expressions
for the radial (transversal) and axial (longitudinal) ionic flow resistances.
To make calculations tracktable, we assume azimuthal and axial symmetry
on the electric potential generated by charged ions and monomer surface
charge densities. We also assume that the radial electrolyte convection
is neglectable. 

In the longitudinal ionic flow resistance calculations, we also assume
that an external perturbation voltage $\triangle V$ is applied between
the monomer ends, which are separated by a distance $\ell$. This
voltage drop generates a uniform axial electric field along the monomer
with magnitude $E_{z}=\frac{\left|\triangle V\right|}{\ell}$, and
consequently, an electro-osmotic (migration) force on the ions in
the electrolyte. Another longitudinal driving force considered in
this approach is the natural convection arising from the movement
of the fluid characterized by the axial velocity profile $v_{z}(r)$.
As a result, the transport law provides the following equation for
the axial electric surface current density \cite{Newmancha1}

\begin{equation}
i_{z}(r)=k(r)E_{z}+v_{z}(r)\rho_{e}(r)\ \ \ \ \ \ \ r\geq r_{\xi}>R\label{eq_49-1}
\end{equation}
This expression depends on the electrolyte conductivity 

\begin{equation}
k(r)=F^{2}\sum_{i}z_{i}^{2}u_{i}c_{i}(r)\label{eq_50-1}
\end{equation}
and the total charge density distribution
\begin{equation}
\rho_{e}(r)=F\sum_{i}z_{i}c_{i}(r)\label{eq_51-1}
\end{equation}

In previous expressions, $r_{\xi}\simeq R$ is the slip velocity position,
$v_{z}(r_{\xi})=0$, $F$ is Faraday's constant and the mobility,
valence, and concentration of ion species $i$ are represented by
$u_{i}$, $z_{i}$, and $c_{i}(r)$, respectively. 

Moreover, we combine Poisson's and the Navier-Stokes' equations to
obtain the following expression for the axial velocity profile 

\begin{equation}
v_{z}(r)=\frac{\epsilon E_{z}}{\mu}\left[\phi(r)-\phi(r_{\xi})\right]\ \ \ \ \ \ \ r\geq r_{\xi}.\label{eq_54-1-1}
\end{equation}
This expression indicates that under the action of a uniform axial
electric field, the velocity profile of the fluid is proportional
to the radial electric potential drop $\phi(r)-\phi(r_{\xi})$. In
the latter equation, $\epsilon=7.0832\cdot10^{-10}\frac{F}{m}$ and
$\mu=0.00089\frac{Kg}{m.s}$ represent the absolute bulk permittivity
and viscosity parameters, respectively. On the other hand, the radial
electric potential $\phi(r)$ is generated by the total charge density
distribution as dictated by the Poisson's equation

\begin{equation}
\rho_{e}(r)=-\frac{\epsilon}{r}\frac{\partial}{\partial r}\left(r\frac{\partial\phi(r)}{\partial r}\right)\label{eq_56-1-1}
\end{equation}

To obtain an analytic solution for the electric potential we use a
Boltzmann distribution and Debye-Hückel (linearized PB) approximation
for the ion density distributions

\begin{equation}
c_{i}(r)=c_{i}^{\infty}\exp\left[-\frac{z_{i}F\phi(r)}{RT}\right]\thickapprox c_{i}^{\infty}\left(1-\frac{z_{i}F\phi(r)}{RT}\right),\qquad\left|\frac{z_{i}F\phi(r)}{RT}\right|\ll1\label{eq_64-1-1-1}
\end{equation}
where $c_{i}^{\infty}$ is the bulk concentration of species $i$
$\left[\frac{mol}{m^{3}}\right]$, $R$ the gas constant, and $T$
the electrolyte temperature. After substitution of eqn (\ref{eq_64-1-1-1})
into eqn (\ref{eq_51-1}), the use of the bulk electroneutrality condition
($\sum_{i}z_{i}c_{i}^{\infty}=0$), and the replacement of the resulting
expression into eqn (\ref{eq_56-1-1}), we obtain
\begin{equation}
\phi(r)=\frac{\sigma\lambda}{\epsilon}\frac{K_{0}\left(\frac{r}{\lambda}\right)}{K_{1}\left(\frac{r_{\xi}}{\lambda}\right)}\qquad r>R\label{eq_75-1-1}
\end{equation}
where $K$ is the modified Bessel function of the second kind, $\lambda$
represents the Debye length $\lambda=\left(\epsilon RT/\left(F^{2}\sum_{i}z_{i}^{2}c_{i}^{\infty}\right)\right)^{\frac{1}{2}}$,
$\sigma$ the monomer surface charge density, $r$ is the radius predicted
by the cylindrical model. Note that the Debye length in eqn (\ref{eq_75-1-1})
plays an important role providing an estimate on the width of the
electrical double layer. 

The approximate analytic solution obtained for the electric potential
$\phi(r)$ is subsequently replaced into eqn (\ref{eq_64-1-1-1}),
(\ref{eq_56-1-1}), (\ref{eq_54-1-1}), and (\ref{eq_50-1}) to get
an analytic solution for the surface current density. This solution
is integrated over the Bjerrum length $\ell_{b}=\frac{e^{2}}{4\pi\varepsilon\varepsilon_{0}k_{b}T}=6.738\mathring{A}$,
e.g. the length scale below which electrostatic correlations are important.
In the latter expressions, $\varepsilon=80$ is the relative bulk
solvent dielectric constant; $\varepsilon_{0}=8.854\cdot10^{-12}\left[\frac{F}{m}\right]$
the vacuum permittivity; $k_{b}=1.381\cdot10^{-23}\left[\frac{J}{K}\right]$
the Boltzmann's constant, and $T$ the temperature in Kelvin degree.
The integration yields the following expression for the total longitudinal
ionic current

\begin{equation}
\frac{I_{l}}{2\pi}=E_{z}\intop_{r_{\xi}}^{\ell_{b}+r_{\xi}}rk(r)dr+\intop_{r_{\xi}}^{\ell_{b}+r_{\xi}}rv(z)\rho_{e}(r)dr\label{eq_55-1-1}
\end{equation}

After performing some algebra and integral calculations using mathematica11.1
software \cite{math} expression \ref{eq_55-1-1} becomes
\begin{multline}
I_{l}=E_{z}\pi\left((\ell_{b}+r_{\xi})^{2}-r_{\xi}^{2}\right)\left\{ k^{\infty}+\triangle k_{l}\right\} =\frac{\left|\Delta V\right|}{\ell}\pi\left((\ell_{b}+r_{\xi})^{2}-r_{\xi}^{2}\right)\left\{ k^{\infty}+\triangle k_{l}\right\} \label{eq_89-1-1}
\end{multline}
where $k^{\infty}$ is the bulk electrolyte conductivity

\begin{equation}
k^{\infty}=F^{2}\sum_{i}z_{i}^{2}u_{i}c_{i}^{\infty},\label{eq_50-1-1-1}
\end{equation}
 $\triangle k_{l}$ the corrections predicted by our approach

\begin{equation}
\triangle k_{l}=-\frac{2F^{3}\sigma\lambda^{2}r_{\xi}\sum_{i}z_{i}^{3}u_{i}c_{i}^{\infty}}{\epsilon RT((\ell_{b}+r_{\xi})^{2}-r_{\xi}^{2})}\left(1-\frac{\left(\ell_{b}+r_{\xi}\right)K_{1}\left(\frac{\ell_{b}+r_{\xi}}{\lambda}\right)}{r_{\xi}K_{1}\left(\frac{r_{\xi}}{\lambda}\right)}\right)+\label{eq:correctionlong}
\end{equation}

\[
\frac{r_{\xi}^{2}\sigma^{2}}{\mu\left((\ell_{b}+r_{\xi})^{2}-r_{\xi}^{2}\right)}G\left(\ell_{b},r_{\xi},\lambda\right)
\]
and $G$ is the following analytic function

\[
G\left(\ell_{b},r_{\xi},\lambda\right)\left[K_{1}\left(\frac{r_{\xi}}{\lambda}\right)\right]^{2}=\left\{ \left(K_{0}\left(\frac{r_{\xi}}{\lambda}\right)^{2}-K_{1}\left(\frac{r_{\xi}}{\lambda}\right)^{2}\right)+2\frac{\lambda}{r_{\xi}}K_{0}\left(\frac{r_{\xi}}{\lambda}\right)K_{1}\left(\frac{r_{\xi}}{\lambda}\right)\right.
\]

\[
\left.-\frac{(\ell_{b}+r_{\xi})^{2}\left(K_{0}\left(\frac{(\ell_{b}+r_{\xi})}{\lambda}\right)^{2}-K_{1}\left(\frac{(\ell_{b}+r_{\xi})}{\lambda}\right)^{2}\right)-2\frac{\lambda(\ell_{b}+r_{\xi})}{r_{\xi}^{2}}K_{0}\left(\frac{(\ell_{b}+r_{\xi})}{\lambda}\right)K_{1}\left(\frac{(\ell_{b}+r_{\xi})}{\lambda}\right)}{r_{\xi}^{2}}\right\} .
\]

Finally, the longitudinal ionic flow resistance $R_{l}$ is calculated
from ohm's law, which relates the axial voltage drop $\Delta V$ and
the total electric current $I$ as $R_{l}=\left|\Delta V/I_{l}\right|$.
As a result, we have

\begin{equation}
R_{l}=\frac{\ell}{\pi\left((\ell_{b}+r_{\xi})^{2}-r_{\xi}^{2}\right)\left|k^{\infty}+\triangle k_{l}\right|}=\frac{\ell}{S_{l}\varrho_{l}},\label{eq_90-1-1}
\end{equation}
where $S_{l}=\pi\left((\ell_{b}+r_{\xi})^{2}-r_{\xi}^{2}\right)$
represents the effective cross section surface area facing perpendicular
to the longitudinal ionic flow and $\varrho_{l}=\left|k^{\infty}+\triangle k_{l}\right|$
is the effective axial ionic conductivity.

A similar approach is used for the transversal ionic flow resistance
calculations. Here we assume that the electro-osmosis generated by
the gradient of the radial electric potential is the only driving
(migration) force dominating the radial surface current density 

\begin{equation}
i_{r}(r)=-F\sum_{i}z_{i}\left(Fz_{i}u_{i}c_{i}(r)\frac{\partial\phi(r)}{\partial r}\right)\label{eq_168}
\end{equation}

Since this expression does not depend on the axial and azimutal coordinates,
the total radial current $I_{r}$ passing from the inner to the outer
layer is obtained by multiplying the radial surface current density
by the lateral monomeric surface layer area at a section $r$ in the
solution, e.g. $I_{r}=i_{r}(r)2\pi\ell r$. \cite{Newmancha1} Since
$I_{r}$ is a constant independent of position, this expression can
be integrated across the electrical double layer to obtain

\begin{equation}
\intop_{r_{\xi}}^{\ell_{b}+r_{\xi}}\frac{I_{r}}{2\pi\ell r}dr=\intop_{r_{\xi}}^{\ell_{b}+r_{\xi}}i_{r}(r)dr=-k^{\infty}\intop_{r_{\xi}}^{\ell_{b}+r_{\xi}}\frac{\partial\phi(r)}{\partial r}dr+\frac{F^{3}}{RT}\sum_{i}z_{i}^{3}u_{i}c_{i}^{\infty}\intop_{r_{\xi}}^{\ell_{b}+r_{\xi}}\phi(r)\frac{\partial\phi(r)}{\partial r}dr\label{eq_176}
\end{equation}

Therefore,

\[
\frac{I_{r}\ln\left(\frac{\ell_{b}+r_{\xi}}{r_{\xi}}\right)}{2\pi\ell}=-k^{\infty}\left[\phi(\ell_{b}+r_{\xi})-\phi(r_{\xi})\right]+\frac{F^{3}}{2RT}\sum_{i}z_{i}^{3}u_{i}c_{i}^{\infty}\left[\phi^{2}(\ell_{b}+r_{\xi})-\phi^{2}(r_{\xi})\right]
\]

After some algebra we obtain a linear dependence between the total
radial current $I_{r}$ and the electric potential drop across the
electrical double layer $\Delta\phi\equiv\phi(r_{\xi})-\phi(\ell_{b}+r_{\xi})$
following the Ohm-like law equation

\[
\frac{I_{r}\ln\left(\frac{\ell_{b}+r_{\xi}}{r_{\xi}}\right)}{2\pi\ell\left[k^{\infty}-\frac{F^{3}}{2RT}\sum_{i}z_{i}^{3}u_{i}c_{i}^{\infty}\left[\phi(\ell_{b}+r_{\xi})+\phi(r_{\xi})\right]\right]}\equiv I_{r}R_{t}=\Delta\phi
\]
where the radial ionic flow resistance is given by

\begin{equation}
R_{t}=\frac{\ln\left(\frac{\ell_{b}+r_{\xi}}{r_{\xi}}\right)}{2\pi\ell\left|k^{\infty}-\frac{F^{3}}{2RT}\sum_{i}z_{i}^{3}u_{i}c_{i}^{\infty}\left[\phi(\ell_{b}+r_{\xi})+\phi(r_{\xi})\right]\right|}=\frac{\ell_{b}}{2\pi\ell r_{\xi}\left|k^{\infty}+\triangle k_{t}\right|}=\frac{\ell_{b}}{S_{t}\varrho_{t}},\label{eq_60}
\end{equation}
In the later expression $S_{t}=2\pi\ell r_{\xi}$ represents the effective
lateral surface area facing perpendicular to the radial ionic flow,
$\varrho_{t}=\left|k^{\infty}+\triangle k_{l}\right|$ is the effective
radial ionic conductivity, and $\triangle k_{t}$ the corrections
to the bulk electrolyte conductivity

\begin{equation}
\triangle k_{t}=k^{\infty}\left\{ \frac{\ell_{b}}{r_{\xi}\ln\left(\frac{\ell_{b}+r_{\xi}}{r_{\xi}}\right)}-1\right\} -\frac{\ell_{b}\sigma\lambda F^{3}\sum_{i}z_{i}^{3}u_{i}c_{i}^{\ell}}{2r_{\xi}RT\epsilon K_{1}\left(\frac{r_{\xi}}{\lambda}\right)\ln\left(\frac{\ell_{b}+r_{\xi}}{r_{\xi}}\right)}\left[K_{0}\left(\frac{\left(\ell_{b}+r_{\xi}\right)}{\lambda}\right)+K_{0}\left(\frac{r_{\xi}}{\lambda}\right)\right]\label{eq:correctiontrans}
\end{equation}

We note that our results recover more oversimplified approximations
for cytoskeleton filaments when these contributions are neglected
,\cite{Tuszynski2004,Sataric2009} namely

\[
R_{l}\simeq R_{l}^{o}=\frac{\ell}{2\pi\ell_{b}r_{\xi}k^{\infty}},\qquad R_{t}\simeq R_{t}^{o}=\frac{\ell_{b}}{2\pi\ell r_{\xi}k^{\infty}},\qquad\ell_{b}\ll r_{\xi},\quad\sigma=0,\qquad c_{i}^{\infty}\ll1.
\]

Additionally, under the conditions considered in this article, $\ell_{b}\lesssim\lambda$
, which is needed to justify a mean-field theory for the diffuse part
of the electrical double layer.

\subsubsection{Ionic Capacitance}

The Poisson-Boltzmann theory used in the previous calculations is
certainly inaccurate to describe the differential capacitance of electric
double layers in ionic liquids and its correlation with the monomer
surface charge density and electric potential .\cite{JIANG2011153,Kornyshev2007}
In the present article, we use a more sophisticated approach based
on a classical solvation density functional theory (CSDFT).\cite{Marucho2014}
The polyelectrolyte properties of the biomolecule are characterized
by the effective molecular radius $R$ and uniform bare surface charge
density $\sigma$ predicted by the Cong molecular structure model,
whereas the biological environment is represented by an electrolyte
(either alkaline, acid or neutral) solution comprised of ionic species
(characterized by crystal radius, \cite{Marcus88} charge and bulk
concentration), explicit water molecules (characterized by neutral
ions at experimental size ($2.75\textrm{Å}$) and bulk concentration
($55.56M$)) (Fig. 2). The approach was successfully tested on segments
of B-DNAs and its application is extended here to describe the polyelectrolyte
properties of a segment of F-actin (e.g., a monomer).. 

To calculate the capacitance of the cylindrical electrical double
layer we consider a monomer perturbated by changing the pH level (alkaline,
acid and neutral levels) in the solution. These pH changes generate
perturbations on the surface charge density $\sigma$, which are predicted
by the Cong molecular structure model and titration calculations as
explained previously. For each surface charge density value we use
CSDFT to predict the induced changes on the corresponding surface
electrical potential $\psi_{o}$. These two sets of parameter values
can be correlated using a cubic fitting polynomial curve. These curves
are used to calculate the slope analytically to obtain the following
expression for the differential capacitance $C_{d}$ 

\begin{equation}
C_{d}=\frac{d\sigma}{d\psi_{o}}=\widehat{C}_{o}\left(1-2\widehat{b}\psi_{o}+3\widehat{c}\psi_{o}^{2}+\mathcal{O}\left(\psi_{o}^{3}\right)\right)\label{eq_2-2-2}
\end{equation}
Integration of this expression with respect to the electric potential
along the voltage drop $V$ across the electrical double layer leads
to the following expression for the total charge accumulated in the
capacitor \cite{Grahame1947} 

\begin{equation}
Q=2\pi r_{\xi}\ell\widehat{C}_{o}\left(V-\widehat{b}V^{2}\right)=C_{o}\left(V-bV^{2}\right)=VC_{o}\left(1-bV\right)=VC\left(V\right)\label{eq_2-1}
\end{equation}
where $C_{o}$ represents the linear capacitance of the capacitor
and $b$ the theory's parameter characterizing its nonlinear behavior.
Specific values for these parameters in intracellular and in-vitro
conditions are provided below. 

\subsubsection{Self-Inductance}

In reference (24) Faraday's law is used to estimate the self-inductance
$L$ for G-actins which is found to be of the order of pico-Henry

\begin{equation}
L=\frac{\Gamma N^{2}\pi\left(r_{\xi}+\lambda\right)^{2}}{\ell}\simeq1.2\cdot10^{-17}F\label{eq_61}
\end{equation}
where $\Gamma$ and $N$ represent the magnetic permeability and the
number of turns (e.g. how many ions could be lined up along the length
of a monomer), respectively. Here $N\simeq d/\ell$, where $d$ is
the average ion size. 

According to this result, the contribution to the electrical signal
propagation for F-actins is expected to be neglectable, however it
may play an important role in other highly charged filaments. Therefore,
in the present work we include this rough estimate of self-inductance
in the approach. 

\section{Lossy, Nonlinear, Dispersive Transmission Line Model for Microfilaments}

Loan counterions surrounding microfilaments may be transferred from
one charge center (e.g. monomer) to the next, giving rise to locally
restricted excess of partial charges representing the propagation
of electrical impulses along the conducting pathway.\cite{GartzkeC1333}
This kind of charge transfer mechanism between monomers is well known
to generate flow of weakly bound electrons in conducting polymer systems
where the electrical current and propagation velocity are different
in nature. These transmission line models have been successfully demonstrated
to characterize electron currents along conducting polymers, and are
often utilized to describe ionic conductance and electrical impulse
propagation along cytoskeleton filaments. \cite{kolosick1974properties,Sekulic2011,sekulic_2012}
Here we use the sequential arrangement of elementary electric units
introduced in references. \cite{Tuszynski2004} In this case, each
unit represents a single G-actin characterized by the capacitor, resistances
and self-inductance described in the previous section.

\begin{figure}
\begin{centering}
\includegraphics[scale=0.2]{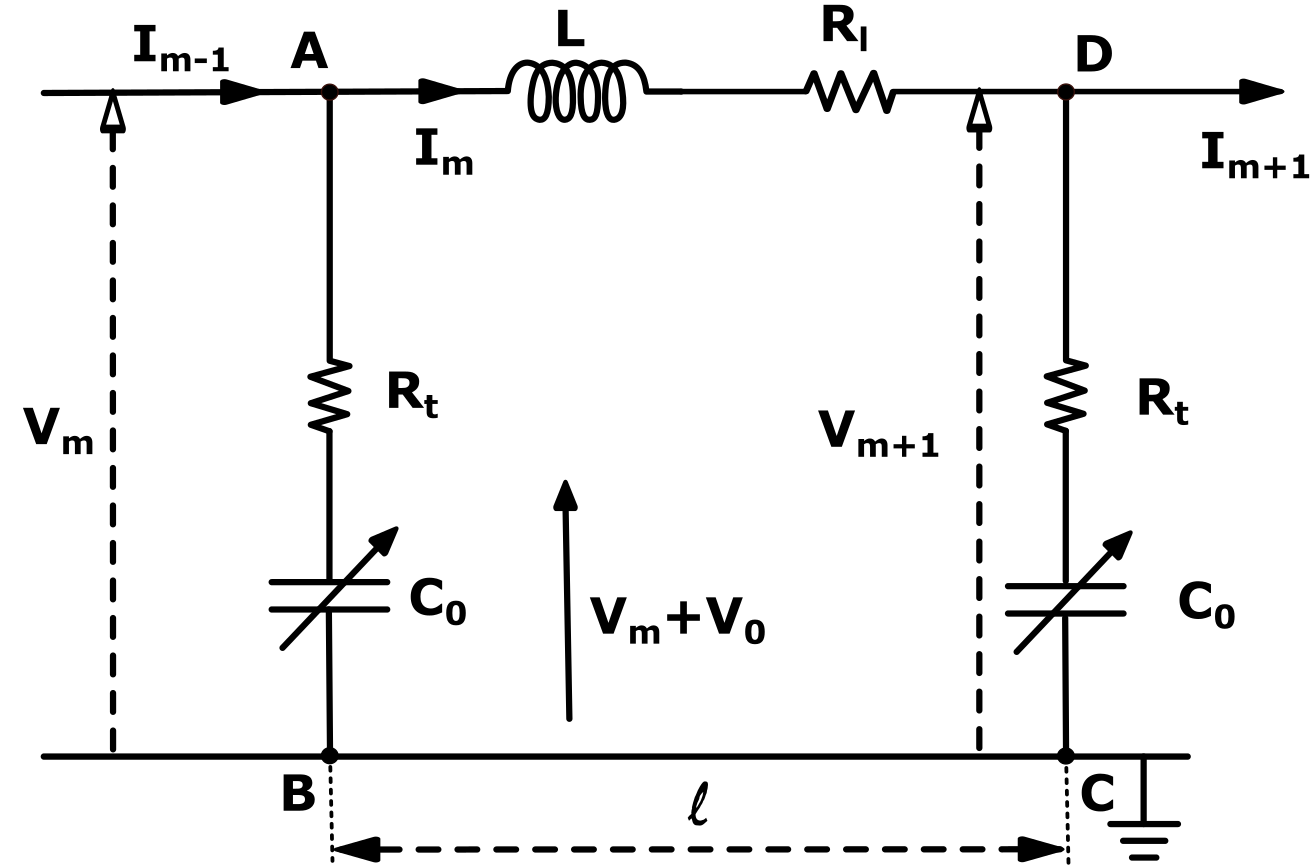}
\par\end{centering}
\centering{}\caption{Effective circuit diagram for the $m^{th}$ monomer\label{fig:Equivalent-electric-cicuit}}
\end{figure}

Application of Kirchoff's laws on the discrete transmission line model
constructed by $N$ elementary cells (monomers) provides a differential
equation for describing the electric potential and ionic current across
cell units. The current conservation law at the node A between the
cells ``$m-1$'' and ``$m$'', provide the following equation
(see Fig. \ref{fig:Equivalent-electric-cicuit})

\begin{equation}
I_{m-1}-I_{m}=\frac{\partial Q_{m}}{\partial t}\label{eq_1}
\end{equation}
where $\partial Q_{m}/\partial t$ represents the current across the
capacitor in the cell ``$m$''. By replacing eqn (\ref{eq_2-1})
into eqn (\ref{eq_1}) we have

\begin{equation}
I_{m-1}-I_{m}=C_{o}\left(\frac{\partial V_{m}}{\partial t}-2bV_{m}\frac{\partial V_{m}}{\partial t}\right)\label{eq_3}
\end{equation}
On the other hand, Kirchhoff's voltage law along the circuit ABCD
generates the following expression (see Fig. \ref{fig:Equivalent-electric-cicuit})

\begin{equation}
v_{m}-v_{m+1}=L\frac{\partial I_{m}}{\partial t}+I_{m}R_{l}\label{eq_4}
\end{equation}
where $v_{m}$ is given by

\begin{equation}
v_{m}=R_{t}\left(I_{m-1}-I_{m}\right)+V_{o}+V_{m}\label{eq_5}
\end{equation}
 and $V_{o}$ represents a constant DC bias electric potential. Further
substitution of eqn (\ref{eq_5}) into eqn (\ref{eq_4}) yields

\begin{equation}
L\frac{\partial I_{m}}{\partial t}+I_{m}R_{l}=R_{t}\left(I_{m-1}-2I_{m}+I_{m+1}\right)+\left(V_{m}-V_{m+1}\right)\label{eq_6}
\end{equation}
The later equation can be written in terms of the characteristic impedance
of the electrical circuit unit $Z$ and the new function $U_{m}(t)$,
where as usual:\cite{Sekulic2011} $Z^{-1/2}U_{m}=I_{m}$ and $Z^{1/2}U_{m}=V_{m}$.
In this article, the characteristic impedance is estimated as follows

\begin{equation}
Z\simeq\sqrt{R_{equiv}^{2}+X_{equiv}^{2}}\label{eq_186}
\end{equation}
where $R_{equiv}=R_{l}+R_{t}$, $X_{equiv}=\frac{T_{o}^{G-actin}}{2\pi C_{o}}$,
and $T_{o}^{G-actin}$ is a parameter characterizing the electrical
circuit unit time scale. 

These expressions, when replaced into eqn (\ref{eq_3}) and (\ref{eq_6}),
lead to the following coupled equations 

\begin{equation}
U_{m-1}-U_{m}=C_{0}Z\left(\frac{\partial U_{m}}{\partial t}-2bZ^{1/2}U_{m}\frac{\partial U_{m}}{\partial t}\right)\label{eq_9}
\end{equation}

\begin{equation}
U_{m}-U_{m+1}=Z^{-1}\left[L\frac{\partial U_{m}}{\partial t}+U_{m}R_{l}-R_{t}\left(U_{m+1}-2U_{m}+U_{m-1}\right)\right]\label{eq_10}
\end{equation}
Since F-actins are a filamentous form of its subunits g-actin, we
have that the individual monomer length $l$ is much smaller than
the filament length $(N-1)l$. As a result, we can approximate eqn
(\ref{eq_9}) and (\ref{eq_10}) as the voltage $V_{m}$ and current
$I_{m}$ travel down the actin filament by moving from one adjacent
circuit modeled monomer to the next. Using the continuum approximation
$U_{m}(t)\backsimeq U(x,t)$ with a Taylor series in terms of the
parameter $\ell$ we obtain the following expansion for $U_{m\pm1}(t)$ 

\begin{equation}
U_{m\pm1}=U(x\pm\ell,t)\backsimeq U\pm\ell\frac{\partial U}{\partial x}+\frac{\ell^{2}}{2}\frac{\partial^{2}U}{\partial x^{2}}\pm\frac{\ell^{3}}{3!}\frac{\partial^{3}U}{\partial x^{3}}\label{eq_11}
\end{equation}
Consequently,

\begin{equation}
U_{m+1}-2U_{m}+U_{m-1}=\ell^{2}\frac{\partial^{2}U}{\partial x^{2}}\label{eq_13}
\end{equation}

\begin{equation}
U_{m-1}-U_{m+1}=-2\ell\frac{\partial U}{\partial x}-\frac{\ell^{3}}{3}\frac{\partial^{3}U}{\partial x^{3}}\label{eq_17}
\end{equation}
By summing eqn (\ref{eq_9}), (\ref{eq_10}), and using eqn (\ref{eq_13})
and (\ref{eq_17}) we have

\begin{multline}
Z^{-1}\left[L\frac{\partial U}{\partial t}+UR_{l}-R_{t}\left(\ell^{2}\frac{\partial^{2}U}{\partial x^{2}}\right)\right]\\
+C_{0}\left(Z\frac{\partial U}{\partial t}-2bZ^{3/2}U\frac{\partial U}{\partial t}\right)=-2\ell\frac{\partial U}{\partial x}-\frac{\ell^{3}}{3}\frac{\partial^{3}U}{\partial x^{3}}\label{eq_18}
\end{multline}
The later equation can be conveniently rewritten as follows

\begin{multline}
\left[\frac{L}{Z}+C_{0}Z\right]\frac{\partial U}{\partial t}+\frac{R_{l}}{Z}U-\frac{R_{t}\ell^{2}}{Z}\frac{\partial^{2}U}{\partial x^{2}}\\
-2bZ^{3/2}C_{0}U\frac{\partial U}{\partial t}+2\ell\frac{\partial U}{\partial x}+\frac{\ell^{3}}{3}\frac{\partial^{3}U}{\partial x^{3}}=0\label{eq_19}
\end{multline}

The master equation (\ref{eq_19}) can be solved for $U(x,t)=U(x(\xi,\tau),t(\tau))=U(\xi,\tau)$
in terms of dimensionless variables

\begin{equation}
\xi=\frac{x}{\beta}-\frac{t}{\alpha},\qquad\tau=\frac{t}{24\alpha},\qquad\textrm{where}\qquad\alpha=\frac{L}{Z}+C_{0}Z>0,\quad\textrm{and}\quad\beta=2\ell,\label{eq_20}
\end{equation}

After some manipulations and using the following relationships $\frac{\partial}{\partial t}=\frac{1}{\alpha}(\frac{\partial}{24\partial\tau}-\frac{\partial}{\partial\xi})$,
$\frac{\partial}{\partial x}=\frac{1}{\beta}\frac{\partial}{\partial\xi}$,
$\frac{\partial^{2}}{\partial x^{2}}=\frac{1}{\beta^{2}}\frac{\partial^{2}}{\partial\xi^{2}}$
and $\frac{\partial^{3}}{\partial x^{3}}=\frac{1}{\beta^{3}}\frac{\partial^{3}}{\partial\xi^{3}}$,
we have

\begin{equation}
\frac{\gamma\partial U}{24\partial\tau}+\frac{\gamma R_{l}}{Z}U-\frac{\gamma R_{t}}{4Z}\left(\frac{\partial^{2}U}{\partial\xi^{2}}\right)+6U\left(\frac{\partial U}{\partial\xi}-\frac{\partial U}{\partial\tau}\right)+\frac{\gamma}{24}\frac{\partial^{3}U}{\partial\xi^{3}}=0\label{eq_27}
\end{equation}
where $\gamma=\frac{3\alpha}{bZ^{3/2}C_{0}}$. Since we look for slow
changes on the time evolution in the electrical impulse solution of
eqn (\ref{eq_27}), $\frac{1}{24}\frac{\partial U}{\partial\tau}\ll\frac{\partial U}{\partial\xi}$,
the time rate in the electrical impulse is considered to be much slower
than on the traveling variable. As a result, an analytic solution
of eqn (\ref{eq_27}) can be obtained by performing the change, $U=-\frac{\gamma}{24}W$
and defining new parameters $\mu_{2}=\frac{6R_{t}}{Z}$, $\mu_{3}=\frac{24R_{l}}{Z}$.
Accordingly, eqn (\ref{eq_27}) becomes the well-known perturbated
Korteweg-de Vries (pKdV) differential equation \cite{Novikov1984,Ablowitz1981}

\begin{equation}
\frac{\partial W}{\partial\tau}-6W\frac{\partial W}{\partial\xi}+\frac{\partial^{3}W}{\partial\xi^{3}}=\mu_{2}\frac{\partial^{2}W}{\partial\xi^{2}}-\mu_{3}W\equiv P\left(W\right)\label{eq_31}
\end{equation}
The left and right sides in the later equation represent the regular
KdV equation \cite{KARPMAN1977307} and the corresponding perturbation,
respectively. The first term to the left resembles the time dependent
term in Fick\textquoteright s diffusion law, whereas the second and
third terms represent the non-linearity and dispersive contributions
arising from the condensed ion cloud in the electrical double layer
and the diffusive spreading of ions along the microfilament, respectively.
On the other hand, the first and second terms to the right represent
the dissipation and damping perturbations, respectively. 

Equation (\ref{eq_31}) appears in a variety of systems \cite{Jawada2014SolitionSO}
describing the propagation of electrical solitons in a nonlinear dispersive
transmission line in the form of localized voltage waves. An initial
pulse $W\left(\xi,0\right)$ in the transmission line may decay into
a sequence of solitons and a tail. In this work, we will consider
single soliton solutions only. In doing so, we assume that the perturbation
is so small that it has a negligible influence on the soliton formation.
Therefore, the perturbation will manifest itself by affecting the
soliton only after an extended amount of time from its origination.
Thus, the mutual interaction between solitons becomes unimportant
when the soliton movement is of the order of its length. In what follows
we consider this problem in the first approximation. In this case,
the solution of equation (\ref{eq_31}) for non-dissipative systems
$\left(R_{l}=R_{t}=0\right)$ represents a non perturbated pulse soliton
of the regular KdV equation (\cite{Karpman77}) 

\begin{equation}
W_{np}(\xi,\tau)=-2\varOmega_{0}^{2}sech^{2}\left[\varOmega_{0}\left(\xi-4\varOmega_{0}^{2}\tau\right)\right]\label{eq_33}
\end{equation}
with dimensionless constant voltage amplitude $2\varOmega_{0}^{2}$
and propagation velocity $4\varOmega_{0}^{2}$. Solitary-wave solutions
that propagate without changing form may also be expected due to a
balance between non-linearity and dispersion \textcolor{black}{(e.g.
$\left|P(W)\right|\simeq0$), hence} requiring that the effect of
the perturbing terms on the shape of the soliton cancel each other
out.\textcolor{black}{{} \cite{Allen2000}} Otherwise\textcolor{black}{,}
when $\mu_{2}$ and/or $\mu_{3}$ are not zero, equation (\ref{eq_33})
is no longer the solution of the perturbed KdV equation. \cite{Karpman77,KARPMAN1977307,Karpman1979,Mas80} 

In this analysis, we look for an analytic solution of eqn (\ref{eq_33})
in the framework of the perturbation theory on the basis of the adiabatic
approximation. \cite{Karpman77} In that case, the solution is a soliton
pulse $W(\xi,\tau)$ in the form :

\begin{equation}
W(\xi,\tau)=-2\left[\varOmega\left(\tau\right)\right]^{2}sech^{2}\left[\varOmega\left(\tau\right)\left(\xi-\eta\left(\tau\right)\right)\right]\label{eq_34}
\end{equation}
where $\varOmega\left(\tau\right)$ and $\eta\left(\tau\right)$ satisfy
the following equations 

\begin{equation}
\frac{d\varOmega}{d\tau}=-\frac{1}{4\varOmega}\intop_{-\infty}^{\infty}P(W)sech^{2}zdz,\label{eq_35}
\end{equation}

\begin{equation}
\frac{d\eta}{d\tau}=4\varOmega^{2}-\frac{1}{4\varOmega^{3}}\intop_{-\infty}^{\infty}P(W)\left[z+\frac{1}{2}sinh(2z)\right]sech^{2}zdz,\label{eq_36}
\end{equation}
 $z=\varOmega\left(\tau\right)\left(\xi-\eta\left(\tau\right)\right)$
and $P(W)$ is the perturbation term defined in eqn (\ref{eq_31}).
The calculation of the integrals appearing in eqn (\ref{eq_35}) and
(\ref{eq_36}) provide the following analytic solutions

\begin{equation}
\begin{array}{c}
\varOmega\left(\tau\right)=\varOmega_{0}\sqrt{\frac{\exp\left(-\frac{4\tau\mu_{3}}{3}\right)}{1+\frac{4\mu_{2}\varOmega_{0}^{2}}{5\mu_{3}}\left(1-\exp\left(-\frac{4\tau\mu_{3}}{3}\right)\right)}}\equiv\varOmega_{0}M_{1}\left(\tau\right)\\
\simeq\varOmega_{0}\left(1-\frac{2}{15}\left(4\mu_{2}\varOmega_{0}^{2}+5\mu_{3}\right)\tau+\mathcal{O}\left((\frac{4\tau\mu_{3}}{3})^{2}\right)\right)
\end{array}\label{eq_37}
\end{equation}

\begin{multline}
\begin{array}{c}
\\
\eta(\tau)=-\frac{5}{4\mu_{2}}\left[4\mu_{3}\tau+3\log\left[5\mu_{3}\right]\right]\\
-\frac{5}{4\mu_{2}}\left[-3\log\left[-4\varOmega_{0}^{2}\mu_{2}+\exp\left(\frac{4\tau\mu_{3}}{3}\right)\left(4\varOmega_{0}^{2}\mu_{2}+5\mu_{3}\right)\right]\right]\equiv4\varOmega_{0}^{2}\tau M_{2}\left(\tau\right)
\end{array}\\
\simeq4\varOmega_{0}^{2}\tau\left(1-\frac{2}{15}\left(4\mu_{2}\varOmega_{0}^{2}+5\mu_{3}\right)\tau+\mathcal{O}\left((\frac{4\tau\mu_{3}}{3})^{2}\right)\right)\label{eq_38}
\end{multline}

The validation of these exppressions is discussed in Appendix A, where
we provide a comparison between the exact numerical and our the approximate
analytic solution.

Note that $V=-\frac{\gamma\sqrt{Z}}{24}W$, therefore the unperturbated
dimensionless amplitude $\varOmega_{0}^{2}$ is linearly proportional
to the external voltage input $V_{inp}$

\begin{equation}
\varOmega_{0}^{2}=24V_{inp}/(Z^{1/2}\left|\gamma\right|)\label{eq:voltageimpact}
\end{equation}

It is worth mentioning that eqn (\ref{eq_37}), (\ref{eq_38}) and
(\ref{eq_39}) describe the evolution of one soliton in the presence
of a perturbation characterized by the amplitudes $M_{1}\left(\tau\right)$
and $M_{2}\left(\tau\right)$ which are expected to yield a slow change
on the soliton parameters. \cite{Karpman77} Therefore, the expansions
appearing in eqn (\ref{eq_37}) and (\ref{eq_38}) provide a good
estimation of the characteristic soliton travel time (in seconds)
$T_{0}^{soliton}=\frac{360\alpha}{2\left(4\mu_{2}\varOmega_{0}^{2}+5\mu_{3}\right)}$.
This time should be, in principle, larger than the characteristic
ion flow time, such that $T_{0}^{soliton}\gtrsim T_{0}^{G-actin}=2\pi C_{o}\sqrt{Z^{2}-\left(R_{l}+R_{t}\right)^{2}}$.
Additionally, for the electrolyte conditions and models considered
in the present article, we have $\varOmega_{0}^{2}\lesssim1$ , $L/Z\ll1$
and $R_{t}/R_{l}\ll1$ which yield the following approximate implicit
equation for the impedance

\[
\frac{360\alpha}{2\left(4\mu_{2}\varOmega_{0}^{2}+5\mu_{3}\right)}\simeq\left(\frac{3ZC_{o}}{16}\frac{Z}{R_{l}}\right)\gtrsim2\pi C_{o}\sqrt{Z^{2}-\left(R_{l}+R_{t}\right)^{2}}\simeq2\pi C_{o}\sqrt{Z^{2}-R_{l}^{2}}
\]
with solution $Z\gtrsim25.1128R_{l}$ and $T_{0}^{soliton}\gtrsim50.1858\pi C_{o}R_{l}$. 

Another important soliton characterization is given by the expression
for the kern velocity of the electrical impulse along the filament
$v(t)$ (in units of $m/s$). By rewriting the argument of solution
(\ref{eq_34}) in terms of the original variables we have

\begin{equation}
\varOmega\left(\tau\right)\left(\xi-\eta\left(\tau\right)\right)=\frac{\varOmega\left(\tau\right)}{\beta}\left(x-\left[\frac{t}{\alpha}\beta+\beta\eta\left(\tau\right)\right]\right).\label{eq:}
\end{equation}
which provides the following expressions for the time dependent wave
number $k(t)=\frac{\varOmega\left(t/(24\alpha)\right)}{\beta}$ and
kern propagation velocity

\begin{equation}
v(t)=\left(\frac{\beta}{\alpha}+\beta\frac{d\eta(\tau)}{dt}\right)=\frac{\beta}{\alpha}\left(1+\frac{1}{24}\left.\frac{d\eta(\tau)}{d\tau}\right|_{\tau=t/(24\alpha)}\right)\label{eq:velo}
\end{equation}
where

\textrm{
\begin{equation}
\frac{d\eta}{d\tau}=4\varOmega_{0}^{2}\frac{\exp\left(-\frac{4\tau\mu_{3}}{3}\right)}{1+\frac{4\mu_{2}\varOmega_{0}^{2}}{5\mu_{3}}\left(1-\exp\left(-\frac{4\tau\mu_{3}}{3}\right)\right)}\label{eq_39}
\end{equation}
}

Based on these results, the time average kern velocity of the soliton
in a time interval $\left[0,t_{max}\right]$ can be calculated as
usual

\begin{equation}
v_{av}=\left\langle v(t)\right\rangle =\frac{1}{t_{max}}\int_{0}^{t_{max}}v(t)dt\label{eq:averagevelocity}
\end{equation}

Here $t_{max}$ represents the vanishing time, which in this work
is considered as the time taken by the initial soliton amplitude to
be attenuated $99\%$, namely $\left[\varOmega\left(\frac{t_{max}}{24\alpha}\right)/\varOmega\left(0\right)\right]^{2}=0.01$.

\section{Results and Discussion}

In this section, we use expressions (\ref{eq_90-1-1}), (\ref{eq_60}),
(\ref{eq_2-2-2}), (\ref{eq_61}), (\ref{eq_34}), (\ref{eq_37})
and (\ref{eq_38}) to investigate the impact of different electrolyte
solutions and voltage stimulus on the physicochemical properties of
G-actins and electrical signal propagation along F-actins. We investigate
two electrolyte solutions, one representing an intracellular biological
environment in physiological solution conditions (140mM $K^{+}$,
4mM $Cl^{-}$, 75mM $HPO_{4}^{2-}$, and 012mM $Na^{+}$ at 310 K),
\cite{Lodish2000} whereas the other represents in vitro conditions
\cite{actin_like_cable_cantiello_1993} (0.1M $K^{+}$ and 0.1M $Cl^{-}$
at 298 K). For the voltage stimulus, we consider both $0.05V$ and
$0.15V$ peak voltage inputs in order to simulate the typical electric
potential present in cells and single microfilament experiments.

\subsection{Model's Parameters}

Our calculations on the Debye length reveal the formation of a wider
electrical double layer in the intracellular condition than in-vitro
condition, namely

\[
\lambda=\begin{array}{cc}
6.587\mathring{A} & \textrm{intracellular fluid}\\
9.902\mathring{A} & \textrm{in vitro }
\end{array}
\]

The effective conductivity predicted by our approach relative to the
conventional results reads

\[
\frac{\varrho_{t}}{k^{\infty}}=\begin{array}{cc}
\left|3.0736-2.1750\right|/3.0736=0.292 & \textrm{intracellular fluid}\\
\left|1.4988+0.1512\right|/1.4988=1.101 & \textrm{in vitro }
\end{array}
\]

\[
\frac{\varrho_{l}}{k^{\infty}}=\begin{array}{cc}
\left|3.0736-1.376\right|/3.0736=0.552 & \textrm{intracellular fluid}\\
\left|1.4988+0.776\right|/1.4988=1.518 & \textrm{in vitro }
\end{array}
\]

with corresponding resistances

\[
R_{t}=\begin{array}{cc}
\frac{1}{0.292}R_{t}^{o}=3.4240R_{t}^{o}=8.71M\Omega & \textrm{intracellular fluid}\\
\frac{1}{1.101}R_{t}^{o}=0.908R_{t}^{o}=4.74M\Omega & \textrm{in vitro }
\end{array}
\]

\[
R_{l}=\begin{array}{cc}
\frac{1}{0.552}R_{l}^{o}=1.8116R_{l}^{o}=261.35M\Omega & \textrm{intracellular fluid}\\
\frac{1}{1.518}R_{l}^{o}=0.6568R_{l}^{o}=195.00M\Omega & \textrm{in vitro }
\end{array}
\]

and impedances

\[
Z=\begin{array}{cc}
6.55409\cdot10^{9}\Omega & \textrm{intracellular fluid}\\
4.89680\cdot10^{9}\Omega & \textrm{in vitro }
\end{array},
\]

The results predicted on ion conductivity indicate an increase (decrease)
of the resistances in typical in-vitro (intracellular) environments
compared to the corresponding bulk values. These corrections, given
by expressions (\ref{eq:correctionlong}), display a balance (competition)
between migration and convection forces, which depend in a nontrival
fashion, on the electrical double layer thickness, the Debye length,
the electrochemistry (particles electropheresis mobility, valence,
bulk density and size, solvent viscosity and dielectric permittivity),
as well as the monomer surface chemistry (surface charge density and
size). 

\begin{figure}
\begin{centering}
\includegraphics[scale=0.35]{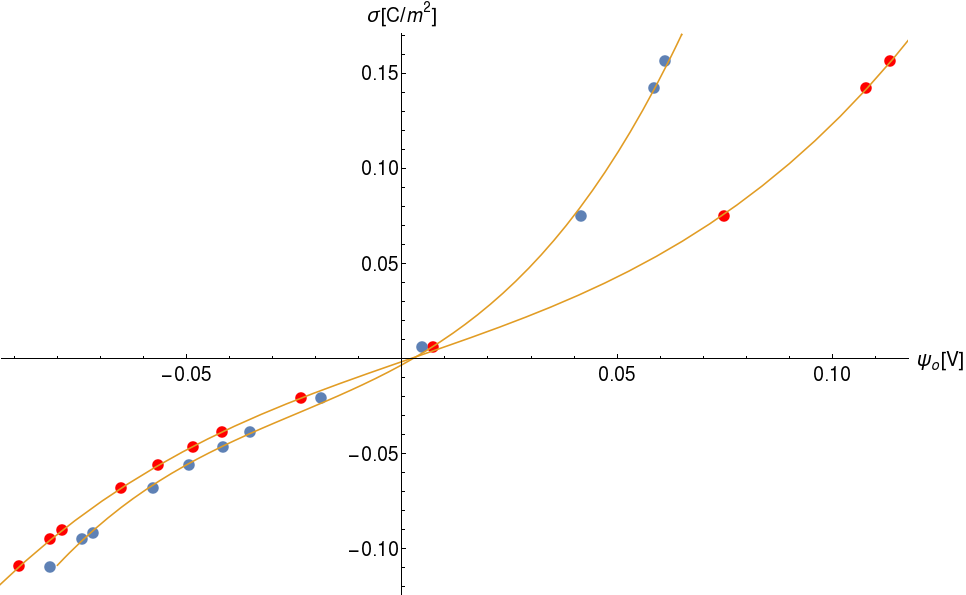}
\par\end{centering}
\caption{\label{fig:fittingcurve}Monomer surface charge density as a function
of the surface electric potential. Blue and red circles represent
the data for intracellular and in-vitro conditions, respectively. }
\end{figure}

Another neoteric result of this work is the prediction of the nonlinear
charge accumulation due to the linear monomeric capacitance behavior,
namely $C(V)=C_{o}\left(1-bV\right)$.. To obtain the numerical values
of the parameters $C_{o}$ and $b$, we correlate the set of surface
charge densities values $\sigma$ predicted by the Cong model with
the set of surface electrical potentials values $\psi_{o}$ predicted
by CSDFT. We use the Fit function provided by mathematica software
\cite{math} to generate a cubic fitting polynomial between these
two parameters as shown in Fig. \ref{fig:fittingcurve}. These curves,
when used to calculate the slope analytically, generate the following
value for $C_{o}$ and $b$

\[
C_{o}=\begin{array}{cc}
1.069\cdot10^{-16}F & \textrm{intracellular fluid}\\
6.739\cdot10^{-17}F & \textrm{in vitro }
\end{array}
\]

\[
b=\begin{array}{cc}
-9.446V^{-1} & \textrm{intracellular fluid}\\
0.4735V^{-1} & \textrm{in vitro }
\end{array}
\]

The results on the capacitor show a remarkable increase in the linear
capacitance in the intracellular condition, which has a high impact
on the monomer's ability to accumulate electric energy in the capacitor.
Additionally, the parameter $b$ is negative for the intracellular
condition, whereas it is positive for the in-vitro condition. Accordingly,
the nonlinearity of the charge accumulated in the capacitor mimics
the behavior of a nMOS varactor in accumulation mode and a diode \cite{Afshari2005}
in our electric circuit unit model for the intracellular and in-vitro
conditions, respectively. We note that the sign of the parameter $b$
also affects the polarization of the transmission line voltage (soliton)
$V=-\frac{\gamma\sqrt{Z}}{24}W$. Certainly, the following caculations

\begin{equation}
\gamma\sqrt{Z}=\frac{3\alpha}{bZ^{3/2}C_{0}}=\frac{3\left(\frac{L}{Z}+C_{0}Z\right)Z^{1/2}}{bZ^{3/2}C_{0}}\simeq\frac{3}{b}=\begin{array}{cc}
-0.3175V & \textrm{intracellular fluid}\\
6.3356V & \textrm{in vitro }
\end{array}\rightarrow V\simeq-\frac{1}{8b}W\label{eq:gammaz}
\end{equation}

predict that the electrical impulse will propagate upright and down
for in-vitro and intracellular conditions, respectively. Another key
role of this parameter arises from eqn (\ref{eq:voltageimpact}) and
(\ref{eq:gammaz}) which predict a linear dependence on the unperturbated
soliton amplitude. Indeed, the substitution of these values for $b$
into eqn (\ref{eq:voltageimpact}) yields $\varOmega_{0}=1.3745$
and $\varOmega_{0}=2.3810$ (intracellular condition) and $\varOmega_{0}=0.30772$
and $\varOmega_{0}=0.5322$ (in-vitro condition), for $0.05V$ and
$0.15V$ voltage inputs, respectively. 

\subsection{Electrical Signal Propagation}

We use expressions \ref{eq_34}, \ref{eq_37} and \ref{eq_38} along
with the numerical values for the parameters obtained previously to
characterize the kern propagation velocity, shape and attenuation
of the electrical impulse along an actin filament under a variety
of conditions.

\begin{figure}
\begin{centering}
\subfloat[Soliton traveling along F-actin in intracellular conditions]{\includegraphics[scale=0.4]{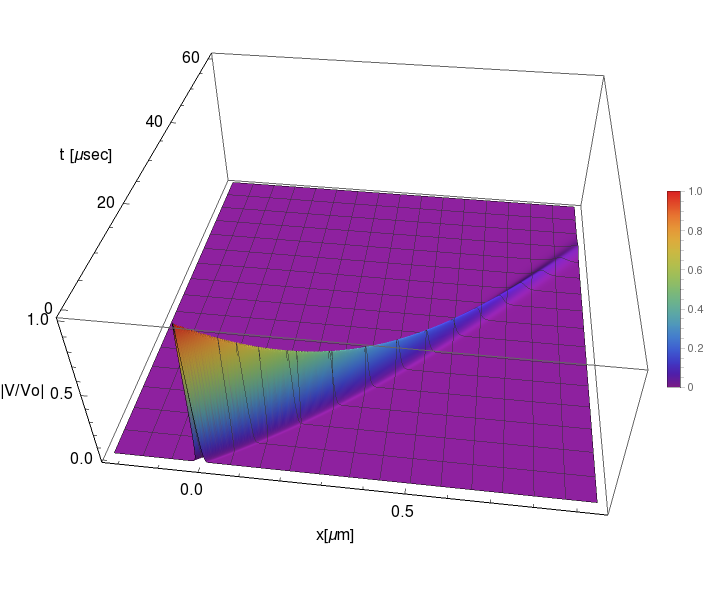}

}\bigskip{}
\par\end{centering}
\begin{centering}
\subfloat[Solition traveling along F-actin in in-vitro conditions]{\includegraphics[scale=0.4]{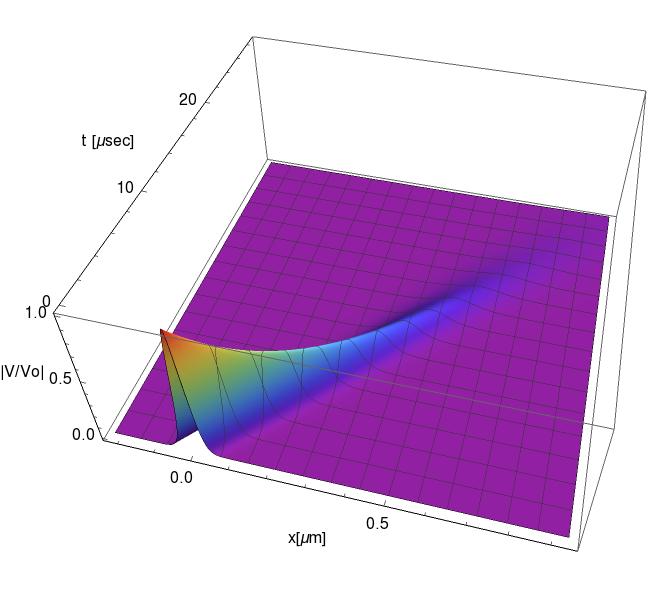}

}
\par\end{centering}
\caption{\label{fig:solitonwaves}Normalized Soliton solution $\left|V\left(x,t\right)/V_{o}\right|=\left|W\left(\frac{x}{\beta}-\frac{t}{\alpha},\frac{t}{24\alpha}\right)/\left(2\Omega_{o}^{2}\right)\right|$
for $0.15V$ input voltage peak. }
\end{figure}

In Fig. \ref{fig:solitonwaves}, we illustrate the propagation of
normalized electrical signals for $0.15V$ voltage peak input for
both intracellular and in-vitro conditions. Our results show similar
soliton range of the order of a micron in both conditions. On the
other hand, the soliton vanishing time is around 2 times longer in
the intracellular condition. As a result, the biological environment
has a high impact on the average kern propagation velocity and attenuation. 

\begin{figure}
\begin{centering}
\subfloat[Soliton profile along F-actin in intracellular conditions. The first,
second, third, fourth and fiftieth peaks (same color) appearing from
left to right correspond to the following snapshots $t=0\mu s$, $t=6\mu s$,
$t=12\mu s$, $t=30\mu s$, and $t=60\mu s$, respectively. ]{\includegraphics[scale=0.35]{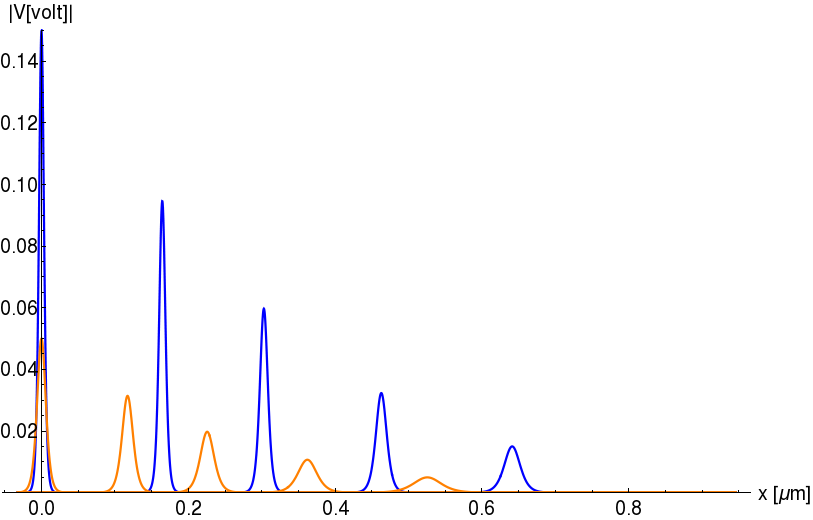}

}
\par\end{centering}
\begin{centering}
\subfloat[Soliton profile along F-actin in in-vitro conditions. The first, second,
third, fourth and fiftieth peaks (same color) appearing from left
to right correspond to the following snapshots $t=0\mu s$, $t=2.8\mu s$,
$t=5.7\mu s$, $t=14.3\mu s$, and $t=28.6\mu s$, respectively. ]{\includegraphics[scale=0.35]{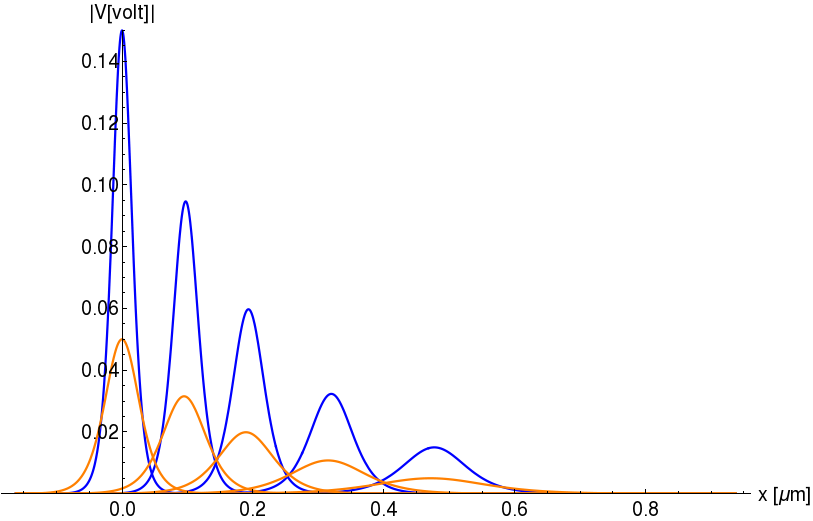}

}
\par\end{centering}
\caption{\label{fig:solitonprofiles}Snapshots of the Normalized soliton solution
$\left|\frac{V\left(x,t\right)}{V_{o}}\right|=\left|\frac{W\left(\frac{x}{\beta}-\frac{t}{\alpha},\frac{t}{24\alpha}\right)}{2\Omega_{o}^{2}}\right|$.
Orange and blue colors represent the electrical signal impulse generated
by $0.05V$ and $0.15V$ input voltage peaks, respectively. }
\end{figure}

\begin{figure}
\begin{centering}
\subfloat[Soliton propagation velocity in intracellular conditions. ]{\includegraphics[scale=0.4]{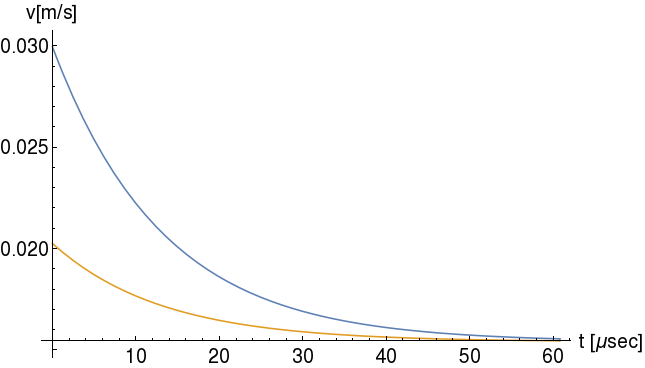}

}\bigskip{}
\par\end{centering}
\begin{centering}
\subfloat[Soliton propagation velocity in in-vitro conditions. ]{\includegraphics[scale=0.4]{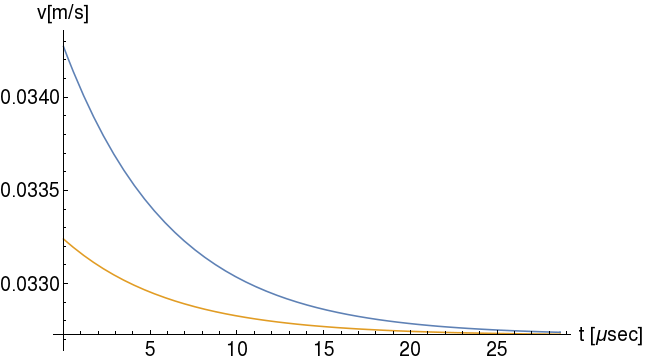}

}
\par\end{centering}
\caption{\label{fig:solitonvelocity}Orange and blue colors represent the propagation
velocity of the electrical signal impulse generated by $0.05V$ and
$0.15V$ input voltage peaks, respectively. }
\end{figure}

\begin{figure}
\begin{centering}
\subfloat[Soliton peak attenuation in intracellular conditions. ]{\includegraphics[scale=0.4]{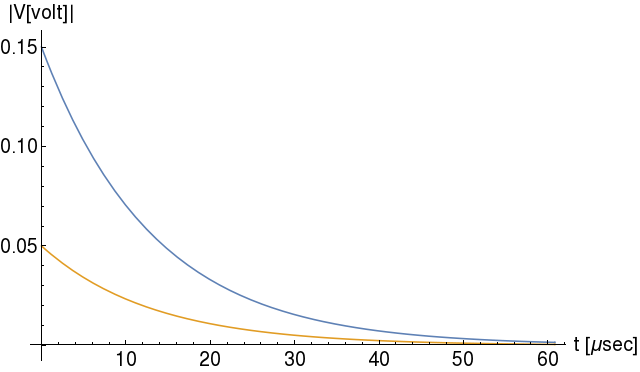}

}\bigskip{}
\par\end{centering}
\begin{centering}
\subfloat[Soliton peak attenuation in in-vitro conditions. ]{\includegraphics[scale=0.4]{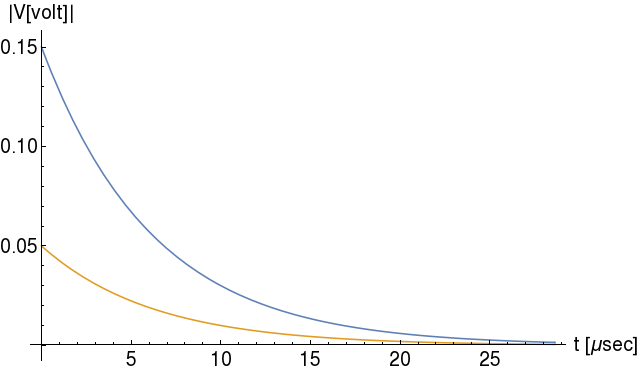}

}
\par\end{centering}
\caption{\label{fig:solitonattenuation}Orange and blue colors represent the
electrical signal impulse amplitude generated by $0.05V$ and $0.15V$
input voltage peaks, respectively. }
\end{figure}

In Fig. \ref{fig:solitonprofiles}, we compare several equitemporal
snapshots of the soliton profile along F-actin for both electrolyte
conditions and voltage peak inputs. In both biological conditions,
the electrical impulse shape is wider for lower voltage inputs. Although,
when comparing to each other, the shape of the soliton is narrower
for the intracellular condition than for the in-vitro condition. The
shift between consecutive blue and orange peak positions shown in
Fig. \ref{fig:solitonprofiles}a) indicates that solitons at higher
voltage input travel faster in the intracellular condition. On the
other hand, Fig. \ref{fig:solitonprofiles}b) implies the voltage
input does not play an important role on the soliton kern velocity
for in-vitro condition. This is in agreement with the results displayed
in Fig. \ref{fig:solitonvelocity} for the soliton kern velocity.
Clearly, the filament is able to sustain the soliton propagation at
almost constant kern velocity in the in-vitro condition (see Fig.
\ref{fig:solitonvelocity}b), namely $v(0)\simeq v_{av}=0.0328m/s$
and $v(0)\simeq v_{av}=0.0331m/s$ for $0.05V$ and $0.15V$ voltage
inputs, respectively. Nevertheless, a different scenario is manifiested
in Fig. \ref{fig:solitonvelocity}a) where the initial kern propagation
velocity for the intracellular condition is around $v(0)=0.03m/s$
and $v(0)=0.02m/s$ for $0.05V$ and $0.15V$ peak voltage inputs,
respectively. The corresponding time averaged kern velocity is much
lower, namely $v_{av}=0.01639m/s$ and $v_{av}=0.011853m/s$. This
indicates a remarkable soliton propagation deceleration caused by
a larger linear capacitance and nonlinear parameter values, higher
longitudinal ionic flow resistance, smaller electrical double layer
thickness, and higher ion asymmetries (size, concentration, electrophoresis
mobility, electrical valence, species number), among other factors.
Additionally, our results demonstrate a higher voltage peak input
that generates a higher time average kern propagation velocity. This
can be understood from eqn (\ref{eq:velo}) and (\ref{eq_39}), which
predict an increasing propagation velocity of the electrical impulse
with $\varOmega_{0}^{2}$, and consequently, with the voltage input
by expression (\ref{eq:voltageimpact}). On the other hand, the time
average velocity comparison between both electrolyte conditions reveal
that solitons from the in-vitro condition travel, on average, faster
than the intracellular condition, while the soliton peak attenuation
is slower in the latter condition. This biological environment impact
on electrical signal propagation is displayed in Fig. \ref{fig:solitonattenuation},
which illustrates the soliton amplitude time evolution. It indicates
similar vanishing time for both voltage peak inputs in each electrolyte
condition. This is a consequence of the neglectable impact of the
amplitude $\varOmega_{0}^{2}$ on the characteristic soliton travel
time $T_{0}^{soliton}=\frac{360\alpha}{2\left(4\mu_{2}\varOmega_{0}^{2}+5\mu_{3}\right)}\simeq\frac{360\alpha}{10\mu_{3}}$.
Moreover, our results reveal a more pronounced, fast soliton attenuation
decay rate at higher voltage input (blue lines in Fig. \ref{fig:solitonattenuation}).
This is caused by higher voltage inputs generating larger values for
$\varOmega_{0}^{2}$, and consequently, larger values for the denominator
in eqn (\ref{eq_37}) for $\varOmega\left(t\right)$.

Overall, our results predict that the propagation of electrical signal
impulses in the form of solitons are possible for a range of electrolyte
solution and voltage stimulus typically present in intracellular and
in-vitro conditions. These predictions are in agreement with avialable
experimental data on single Actin filaments. \cite{actin_like_cable_cantiello_1993,Cantiello1991OSMOTICALLY} 

\section{Conclusions}

In this article, we introduced an innovative multi-scale approach
which accounts for the atomistic details on the protein molecular
structure and biological environment, as well as their impact on electrical
impulses propagating in the form of micron solitons along wild type
Actin filaments. The approach provides a novel, simple, accurate,
approximate analytic expression for the characterization of solitons.
It has been used to determine the effects of electrolyte conditions
and voltage stimulus on the electrical impulse shape, attenuation
and kern propagation velocity. The formulation has been shown to be
capable of accounting for the details on the electrical double layer
thickness and layering formation (ionic and water density distributions),
the electrokinetics (particles electropheresis mobility, valence and
size, solvent viscosity and dielectric permittivity), and the monomer-electrolyte
interface (surface charge density and size) on the ionic electrical
conductivity and capacitance.

Our results reveal a high impact of the electrolyte condition on electrical
conductivity and capacitance in G-actins. The approach predicts wider
electrical double layer, higher electrical conductivity, linear capacitance
and nonlinear accumulation of charge in intracellular condictions,
which play an important role on the electrical signal propagation
along the Actin filament. Additionally, the nonlinearity of the charge
accumulated in the capacitor resembles the behavior of a varactor
and a diode in our electric circuit unit model for the intracellular
and in-vitro conditions, respectively. The approach also predicts
different polarization of the transmission line voltage (soliton).
The electrical impulse propagates upright and down for in-vitro and
intracellular conditions, respectively. 

Our results also show a siginificant influence of the voltage input
on the electrical impulse shape, attenuation and kern propagation
velocity. The filament is able to sustain the soliton propagation
at almost constant kern velocity for the in-vitro condition, but it
displays a remarkable deceleration for the intracellular condition,
with a slower soliton peak attenuation which is more pronounced at
higher voltage input. Solitons are narrower and travel faster at higher
voltage input in the intracellular condition. Whereas the voltage
input does not seem to play an important role on the soliton kern
velocity in the in-vitro condition. On the other hand, the electrical
impulse shape is wider at lower voltage input and the soliton range
of the order of one micron in both electrolyte conditions. Although,
the vanishing time is of around 2 times longer in the intracellular
condition. 

Overall, our results predict that the propagation of electrical signal
impulses in the form of solitons are possible for a range of electrolyte
solutions and voltage stimulus typically present in intracellular
and in-vitro conditions. Our preditions are an improvement on less
recent theories and in good agreement with experimentaly obtained
data for single filaments. One of the most important outcomes of this
approach lies in the ability to determine the impact of molecular
structure conformation (mutations) and physicochemical solution changes
(protonations/deprotonations alterations) often present in pathological
conditions in cytoskeleton filaments. This multi-scale theory may
also be applicable to other highly charged rod-like polyelectrolytes
with relevance in biomedicine and biophysics. \cite{Janmey2014} Currently
we are working along this direction with the ultimate goal of providing
a molecular understanding for how and why age and inheritance conditions
induce dysfunction and malformation in cytoskeleton filaments associated
with a variety of diseases. \cite{dos2008actin}

Our preditions are an improvement on less recent theories and in good
agreement with experimentaly obtained data for single filaments.

\section*{Acknowledgments}

This work was supported by NIH Grant 1SC2GM112578-01. 

\section{Appendix A}

\subsection{Numerical and Analytic Solution Comparison}

\begin{figure}[h]
\begin{centering}
\subfloat[Intracellular conditions. ]{\includegraphics[scale=0.3]{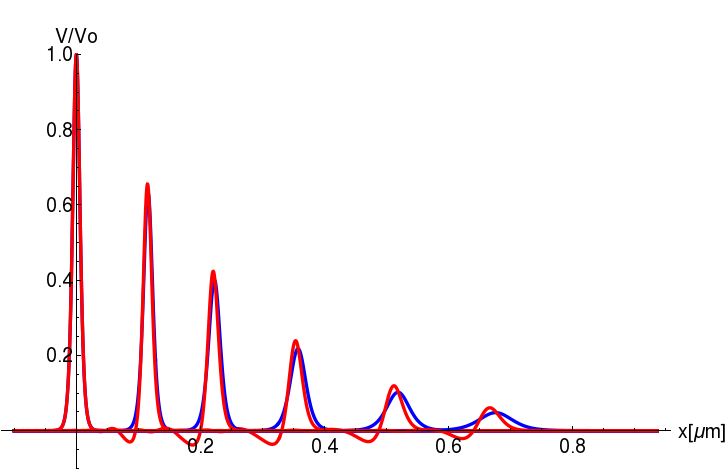}\includegraphics[scale=0.3]{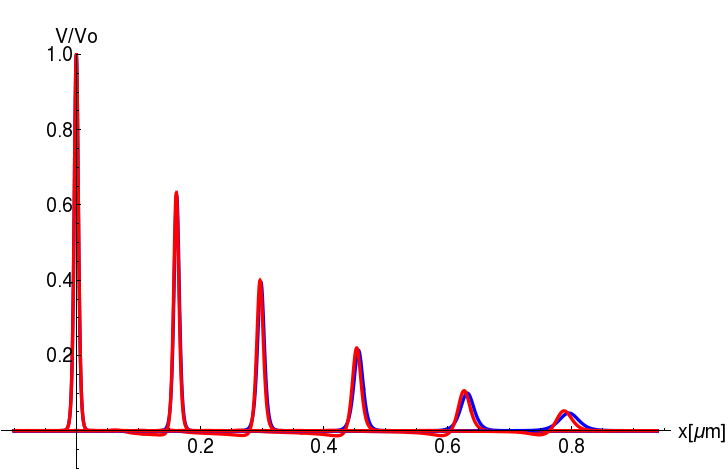}

}\bigskip{}
\par\end{centering}
\begin{centering}
\subfloat[In-vitro conditions ]{\includegraphics[scale=0.3]{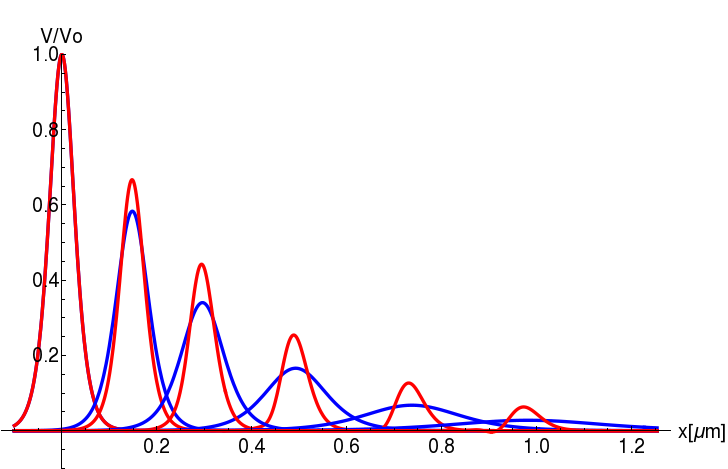}\includegraphics[scale=0.3]{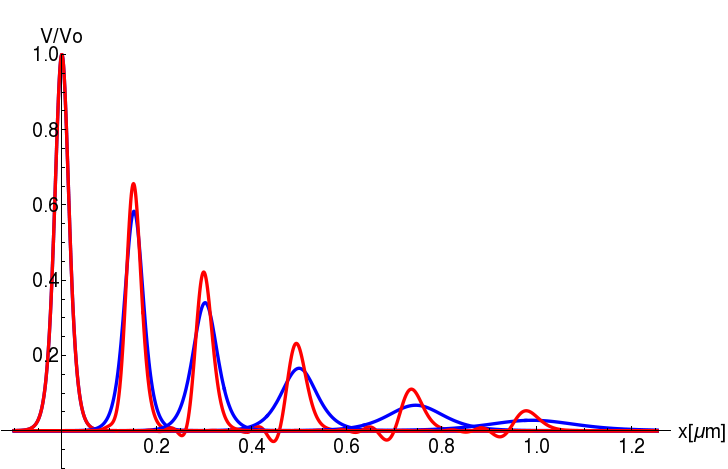}

}
\par\end{centering}
\caption{\label{fig:numandanalcomparison}Comparison between numerical (red
color) and analytic (blue color) solutions. The figures to the left
and right sides correspond to the electrical signal impulse amplitude
generated by $0.05V$ and $0.15V$ input voltage peaks, respectively.
The electrical impulse peaks correspond to the snapshots mentioned
in Fig. \ref{fig:solitonprofiles}}
\end{figure}

We solve equation \ref{eq_31} numerically by using periodic boundary
conditions $W\left(\xi,\tau\right)=W\left(-\xi,\tau\right)$ and a
voltage input signal:

\begin{equation}
W\left(\xi,0\right)=-2\varOmega_{0}^{2}sech^{2}\left[\xi\right]\label{eq_32-1}
\end{equation}
The artificially periodic boundary conditions were imposed to facilitate
the resolution of the partial differential equation \ref{eq_31}.
However, for lengths of the microfilament big enough it does not affect
the solution of the system. \cite{Skogestad_2009} The equation \ref{eq_31}
was solved using the commercial software Mathematica 11.0. \cite{math}
We applied the numerical method of lines algorithm which is a efficient
approach to numerically solve partial differential equations provided
it is an initial value problem. This method discretizes all but one
dimension, then integrates the semi-discrete problem as a system of
Ordinary Differential Equations (ODEs) or Differential-Algebraic Equations
(DAEs). Additionally, we configured some parameters to obtain the
solution. We set the WorkingPrecision (e.g. how many digits of precision
should be maintained in internal computations) to the MachinePrecision
value (double-precision floating-point numbers: \ensuremath{\approx}
16 decimal digits) . The AccuracyGoal and the PrecisionGoal (e.g.
how many effective digits of accuracy and precision, respectively)
were set to a value equal to half the setting for WorkingPrecision.
The InterpolationOrder of the solution (e.g. continuity degree of
the final output) was set to 6 for the $\xi$ variable and 3 for the
$\tau$ variable. The MaxStepFraction (e.g. maximum fraction of the
total range to cover in a single step) was equal to 1/10, the MaxStepSize
(e.g. maximum size of each step) was defined as the inverse of MaxStepFraction
(10) and the MaxSteps (e.g. maximum number of steps to take in generating
a result) was set to 10000. In the case of the NormFunction parameter,
we used an infinity-norm. \cite{math} It is worth mentioning that
the inductance value considered in this work does not affect the numerical
solution obtained for the soliton. 

Fig. \ref{fig:numandanalcomparison} shows the soliton profile comparison
between the numerical and the approximate analytic solution (\ref{eq_34})
for both intracellular and in-vitro conditions, obtaining a good visual
matching over the whole domain. In general, there was a short and
intermediate time evolution where the adiabatic approximation is valid.
Certainly, at longer times the perturbation increases the impact on
the soliton shape and tails. Overall, the peak position and width,
as well as the kern velocity between the numerical and analytic solutions
in intracellular conditions, are in very good agreement. Whereas,
the analytic solution predicts a wider and more attenuated soliton
for in-vitro conditions. Higher order approximations and multisoliton
solutions will be considered in a future work.

\section*{References}

\bibliographystyle{unsrt}
\bibliography{../references2}

\begin{thebibliography}{10}

\bibitem{dos2008actin}
C.~dos Remedios and D.~Chhabra.
\newblock {\em Actin-Binding Proteins and Disease}.
\newblock Protein Reviews. Springer New York, 2008.

\bibitem{woolf2009nanoneuroscience}
N.J. Woolf and A.~Priel.
\newblock {\em Nanoneuroscience: Structural and Functional Roles of the
  Neuronal Cytoskeleton in Health and Disease}.
\newblock Biological and Medical Physics, Biomedical Engineering. Springer
  Berlin Heidelberg, 2009.

\bibitem{Janmey2014}
Paul~A. Janmey, David~R. Slochower, Yu-Hsiu Wang, Qi~Wen, and Andrejs Cebers.
\newblock Polyelectrolyte properties of filamentous biopolymers and their
  consequences in biological fluids.
\newblock {\em Soft Matter}, 10:1439--1449, 2014.

\bibitem{Cantiello1991OSMOTICALLY}
H.F. Cantiello, C.~Patenaude, and K.~Zaner.
\newblock Osmotically induced electrical signals from actin filaments.
\newblock {\em Biophysical Journal}, 59(6):1284 -- 1289, 1991.

\bibitem{actin_like_cable_cantiello_1993}
E.C. Lin and H.F. Cantiello.
\newblock A novel method to study the electrodynamic behavior of actin
  filaments. evidence for cable-like properties of actin.
\newblock {\em Biophysical Journal}, 65(4):1371 -- 1378, 1993.

\bibitem{Goldman1952}
JE~Goldman.
\newblock Immunocytochemical studies of actin localization in the central
  nervous system.
\newblock {\em Journal of Neuroscience}, 3(10):1952--1962, 1983.

\bibitem{LANGE2000561}
Klaus Lange.
\newblock Microvillar ion channels: Cytoskeletal modulation of ion fluxes.
\newblock {\em Journal of Theoretical Biology}, 206(4):561 -- 584, 2000.

\bibitem{GartzkeC1333}
Joachim Gartzke and Klaus Lange.
\newblock Cellular target of weak magnetic fields: ionic conduction along actin
  filaments of microvilli.
\newblock {\em American Journal of Physiology - Cell Physiology},
  283(5):C1333--C1346, 2002.

\bibitem{sundberg2006actin}
Mark Sundberg, Richard Bunk, Nuria Albet-Torres, Anders Kvennefors, Fredrik
  Persson, Lars Montelius, Ian~A Nicholls, Sara Ghatnekar-Nilsson, P{\"a}r
  Omling, Sven T{\aa}gerud, et~al.
\newblock Actin filament guidance on a chip: toward high-throughput assays and
  lab-on-a-chip applications.
\newblock {\em Langmuir}, 22(17):7286--7295, 2006.

\bibitem{Arsenault2007}
Mark~E. Arsenault, Hui Zhao, Prashant~K. Purohit, Yale~E. Goldman, and Haim~H.
  Bau.
\newblock Confinement and manipulation of actin filaments by electric fields.
\newblock {\em Biophysical Journal}, 93(8):L42 -- L44, 2007.

\bibitem{galland2013}
R{\'e}mi Galland, Patrick Leduc, Christophe Gu{\'e}rin, David Peyrade, Laurent
  Blanchoin, and Manuel Th{\'e}ry.
\newblock {Fabrication of three-dimensional electrical connections by means of
  directed actin self-organization.}
\newblock {\em {Nature Materials}}, 12(5):416--21, May 2013.
\newblock A Press release CEA, CNRS, UJF, INRA has been published for this
  publication (February 11th, 2013):
  http://www2.cnrs.fr/presse/communique/2987.htm.

\bibitem{Patolsky2004}
Fernando Patolsky, Yossi Weizmann, and Itamar Willner.
\newblock Actin-based metallic nanowires as bio-nanotransporters.
\newblock {\em Nature Materials}, 3(10):692--5, 10 2004.
\newblock Copyright - Copyright Nature Publishing Group Oct 2004; Last updated
  - 2014-04-30.

\bibitem{Korten2010}
Till Korten, Alf M{\aa}nsson, and Stefan Diez.
\newblock Towards the application of cytoskeletal motor proteins in molecular
  detection and diagnostic devices.
\newblock {\em Current Opinion in Biotechnology}, 21(4):477 -- 488, 2010.

\bibitem{Oosawa_1970}
Fumio Oosawa.
\newblock Counterion fluctuation and dielectric dispersion in linear
  polyelectrolytes.
\newblock {\em Biopolymers}, 9(6):677--688, 1970.

\bibitem{manning_1978}
Gerald~S. Manning.
\newblock The molecular theory of polyelectrolyte solutions with applications
  to the electrostatic properties of polynucleotides.
\newblock {\em Quarterly Reviews of Biophysics}, 11(2):179--246, 1978.

\bibitem{Manning_1969}
G.~S. {Manning}.
\newblock {Limiting Laws and Counterion Condensation in Polyelectrolyte
  Solutions I. Colligative Properties}.
\newblock {\em J. Chem. Phys.}, 51:924--933, August 1969.

\bibitem{Zimm_1986}
Bruno~H. Zimm.
\newblock {\em Use of the Poisson-Boltzmann Equation To Predict Ion
  Condensation Around Polyelectrolytes}, chapter~17, pages 212--215.
\newblock 1986.

\bibitem{kolosick1974properties}
Joseph~A Kolosick, Don~L Landt, HCS Hsuan, and Karle~E Lonngren.
\newblock Properties of solitary waves as observed on a nonlinear dispersive
  transmission line.
\newblock {\em Proceedings of the IEEE}, 62(5):578--581, 1974.

\bibitem{Noguchi1974}
A.~{Noguchi}.
\newblock {Solitons in a nonlinear transmission line}.
\newblock {\em Electronics Communications of Japan}, 57:9 -- 13, Feb 1974.

\bibitem{Lonngren1978127}
Karl~E. Lonngren.
\newblock Observations of solitons on nonlinear dispersive transmission lines.
\newblock In Karl Lonngren and Alwyn Scott, editors, {\em Solutions in Action},
  pages 127 -- 152. Academic Press, 1978.

\bibitem{Novikov1984}
S.~Novikov, S.V. Manakov, L.P. Pitaevskii, and V.E. Zakharov.
\newblock {\em Theory of Solitons: The Inverse Scattering Method}.
\newblock Plenum, New York, 1984.

\bibitem{Bret_Le_Zimm_1984}
Marc~Le Bret and Bruno~H. Zimm.
\newblock Distribution of counterions around a cylindrical polyelectrolyte and
  manning's condensation theory.
\newblock {\em Biopolymers}, 23(2):287--312, 1984.

\bibitem{Newmancha1}
John Newman.
\newblock {\em Electrochemical Systems}, chapter~1.
\newblock Englewood Cliffs, N.J. Prentice-Hall, 1973.

\bibitem{Tuszynski2004}
J.A. Tuszy{\'n}ski, S.~Portet, J.M. Dixon, C.~Luxford, and H.F. Cantiello.
\newblock Ionic wave propagation along actin filaments.
\newblock {\em Biophysical Journal}, 86(4):1890 -- 1903, 2004.

\bibitem{Sataric2009}
M.~V. Satari{\'{c}}, D.~I. Ili{\'{c}}, N.~Ralevi{\'{c}}, and Jack~Adam
  Tuszynski.
\newblock A nonlinear model of ionic wave propagation along microtubules.
\newblock {\em European Biophysics Journal}, 38(5):637--647, Jun 2009.

\bibitem{Sekulic2011}
D.~L. Sekuli{\'{c}}, B.~M. Satari{\'{c}}, J.~A. Tuszy{\'n}ski, and M.~V.
  Satari{\'{c}}.
\newblock Nonlinear ionic pulses along microtubules.
\newblock {\em The European Physical Journal E}, 34(5):49, 2011.

\bibitem{Sekulic_MT_2015}
D.~L. "Sekuli{\'{c}} and B.~M. Satari{\'{c}}.
\newblock An improved nanoscale transmission line model of microtubule: The
  effect of nonlinearity on the propagation of electrical signals.
\newblock {\em Facta Universitatis, Series: Electronics and Energetics},
  28(1):133--142, 2015.

\bibitem{Kornyshev2007}
Alexei~A. Kornyshev.
\newblock Double-layer in ionic liquids: Paradigm change?
\newblock {\em The Journal of Physical Chemistry B}, 111(20):5545 -- 5557,
  2007.

\bibitem{JIANG2011153}
De~en~Jiang, Dong Meng, and Jianzhong Wu.
\newblock Density functional theory for differential capacitance of planar
  electric double layers in ionic liquids.
\newblock {\em Chemical Physics Letters}, 504(4):153 -- 158, 2011.

\bibitem{Lamperski2015}
Stanislaw Lamperski, Monika Pluciennik, and Christopher~W. Outhwaite.
\newblock The planar electric double layer capacitance for the solvent
  primitive model electrolyte.
\newblock {\em Phys. Chem. Chem. Phys.}, 17:928--932, 2015.

\bibitem{Warshavsky2016}
Vadim Warshavsky and Marcelo Marucho.
\newblock Polar-solvation classical density-functional theory for electrolyte
  aqueous solutions near a wall.
\newblock {\em Physical Review E - Statistical, Nonlinear, and Soft Matter
  Physics}, 93(4), 4 2016.

\bibitem{CONG2008331}
Yao Cong, Maya Topf, Andrej Sali, Paul Matsudaira, Matthew Dougherty, Wah Chiu,
  and Michael~F. Schmid.
\newblock Crystallographic conformers of actin in a biologically active bundle
  of filaments.
\newblock {\em Journal of Molecular Biology}, 375(2):331 -- 336, 2008.

\bibitem{Newmancha9}
John Newman.
\newblock {\em Electrochemical Systems}, chapter~9.
\newblock Englewood Cliffs, N.J. Prentice-Hall, 1973.

\bibitem{Marucho2014}
Zaven Ovanesyan, Bharat Medasani, Marcia~O Fenley, Guillermo~Iv{\'a}n
  Guerrero-Garc{\'\i}a, M{\'o}nica Olvera de~la Cruz, and Marcelo Marucho.
\newblock Excluded volume and ion-ion correlation effects on the ionic
  atmosphere around b-dna: Theory, simulations, and experiments.
\newblock {\em The Journal of chemical physics}, 141(22):225103, 2014.

\bibitem{medasani2014ionic}
Bharat Medasani, Zaven Ovanesyan, Dennis~G Thomas, Maria~L Sushko, and Marcelo
  Marucho.
\newblock Ionic asymmetry and solvent excluded volume effects on spherical
  electric double layers: A density functional approach.
\newblock {\em The Journal of chemical physics}, 140(20):204510, 2014.

\bibitem{Hunley2017}
Christian Hunley and Marcelo Marucho.
\newblock Electrical double layer properties of spherical oxide nanoparticles.
\newblock {\em Phys. Chem. Chem. Phys.}, 19:5396--5404, 2017.

\bibitem{Marucho2016}
Zaven Ovanesyan, Amal Aljzmi, Manal Almusaynid, Asrar Khan, Esteban Valderrama,
  Kelly~L. Nash, and Marcelo Marucho.
\newblock Ion-ion correlation, solvent excluded volume and ph effects on
  physicochemical properties of spherical oxide nanoparticles.
\newblock {\em Journal of Colloid and Interface Science}, 462(Supplement C):325
  -- 333, 2016.

\bibitem{Dolinsky2004}
Todd~J. Dolinsky, Jens~E. Nielsen, J.~Andrew McCammon, and Nathan~A. Baker.
\newblock Pdb2pqr: an automated pipeline for the setup of poisson-boltzmann
  electrostatics calculations.
\newblock {\em Nucleic Acids Research}, 32(suppl 2):W665 -- W667, 2004.

\bibitem{3vvoss2010}
Neil~R. Voss and Mark Gerstein.
\newblock 3v: cavity, channel and cleft volume calculator and extractor.
\newblock {\em Nucleic Acids Research}, 38(suppl 2):W555--W562, 2010.

\bibitem{math}
{Wolfram Research, Inc.}
\newblock {\em Mathematica Version 11.0}, 2016.

\bibitem{Marcus88}
Yizhak Marcus.
\newblock Ionic radii in aqueous solutions.
\newblock {\em Chemical Reviews}, 88(8):1475--1498, 1988.

\bibitem{Grahame1947}
David~C. Grahame.
\newblock The electrical double layer and the theory of electrocapillarity.
\newblock {\em Chemical Reviews}, 41(3):441--501, 1947.
\newblock PMID: 18895519.

\bibitem{sekulic_2012}
D.~L. Sekuli{\'{c}} and M.~B. Zivanov.
\newblock Computational study on soliton-like pulses in the nonlinear rlc
  transmission lines.
\newblock In {\em 2012 Proceedings of the 35th International Convention MIPRO},
  pages 228--232, May 2012.

\bibitem{Ablowitz1981}
M.~Ablowitz and H.~Segur.
\newblock {\em Solitons and the Inverse Scattering Transform}.
\newblock Society for Industrial and Applied Mathematics, 1981.

\bibitem{KARPMAN1977307}
V.I. Karpman and E.M. Maslov.
\newblock A perturbation theory for the korteweg-de vries equation.
\newblock {\em Physics Letters A}, 60(4):307 -- 308, 1977.

\bibitem{Jawada2014SolitionSO}
A.~J.~Mohamad Jawada.
\newblock Solition solutions of a few nonlinear wave equations in engineering
  sciences.
\newblock 2014.

\bibitem{Karpman77}
V.~I. Karpman and E.~M. Maslov.
\newblock Perturbation theory for solitons.
\newblock {\em Zh. Eksp. Teor. Fiz.}, 73:537--, August 1977.

\bibitem{Allen2000}
M.~A. Allen and G.~Rowlands.
\newblock A solitary-wave solution to a perturbed kdv equation.
\newblock {\em Journal of Plasma Physics}, 64(4):475--480, 2000.

\bibitem{Karpman1979}
V~I Karpman.
\newblock Soliton evolution in the presence of perturbation.
\newblock {\em Physica Scripta}, 20(3-4):462, 1979.

\bibitem{Mas80}
E.~M. Maslov.
\newblock Perturbation theory for solitons in the second approximation.
\newblock {\em Theoretical and Mathematical Physics}, 42:237--245, 1980.

\bibitem{Lodish2000}
H.~Lodish, A.~Berk, SL. Zipursky, H.~Lodish, P.~Matsudaira, D.~Baltimore, and
  J.~Darnell.
\newblock {\em Molecular Cell Biology. 4th edition}, chapter Section 15.4,
  Intracellular Ion Environment and Membrane Electric Potential.
\newblock New York: W. H. Freeman, 2000.

\bibitem{Afshari2005}
E.~Afshari and A.~Hajimiri.
\newblock Nonlinear transmission lines for pulse shaping in silicon.
\newblock {\em IEEE Journal of Solid-State Circuits}, 40(3):744--752, March
  2005.

\bibitem{Skogestad_2009}
Jan~Ole Skogestad and Henrik Kalisch.
\newblock A boundary value problem for the kdv equation: Comparison of
  finite-difference and chebyshev methods.
\newblock {\em Mathematics and Computers in Simulation}, 80(1):151 -- 163,
  2009.
\newblock Nonlinear Waves: Computation and Theory \{VII\}.

\end{thebibliography}

\end{document}